\documentclass[lettersize,journal]{IEEEtran}
\usepackage{amsmath,amsfonts}
\usepackage{algorithmic}
\usepackage{array}
\usepackage[caption=false,font=normalsize,labelfont=sf,textfont=sf]{subfig}
\usepackage{textcomp}
\usepackage{stfloats}
\usepackage{url}
\usepackage{verbatim}
\usepackage{graphicx}
\def\BibTeX{{\rm B\kern-.05em{\sc i\kern-.025em b}\kern-.08em
    T\kern-.1667em\lower.7ex\hbox{E}\kern-.125emX}}
\usepackage{balance}

\usepackage[utf8]{inputenc} 
\usepackage[T1]{fontenc}    
\usepackage{tikz,xcolor}
\definecolor{citecolor}{RGB}{39,125,154}
\usepackage[colorlinks,linkcolor=blue,citecolor=citecolor]{hyperref}
\usepackage{url}            
\usepackage{booktabs}       
\usepackage{amsfonts}       
\usepackage{nicefrac}       
\usepackage{microtype}      
\usepackage{xcolor}         
\usepackage{subfloat}
\usepackage{graphicx}
\usepackage{multirow}
\usepackage{amsmath}
\usepackage{amsthm}
\usepackage{balance}

\usepackage{graphicx}
\usepackage{multirow}
\usepackage{xspace}
\usepackage{amsmath}
\usepackage{enumitem}
\usepackage[linesnumbered,ruled]{algorithm2e}
\usepackage{wrapfig}

\theoremstyle{definition}
\newtheorem{definition}{Definition}[section]

\newcommand{\ourmethod}{PCKD\xspace}

\begin{document}

\title{Preference-Consistent Knowledge Distillation for Recommender System}

\author{Zhangchi Zhu and Wei Zhang,~\IEEEmembership{Member,~IEEE}
\thanks{This work was supported in part by the National Natural Science Foundation of China under Grant 62072182, Grant 92270119, and the Key Laboratory of Advanced Theory and Application in Statistics and Data Science, Ministry of Education.}
\thanks{Zhangchi Zhu and Wei Zhang are with the School of Computer Science and Technology, East China Normal University, Shanghai, China. (email: zczhu@stu.ecnu.edu.cn, zhangwei.thu2011@gmail.com). (Corresponding author: Wei Zhang).}
\thanks{Manuscript received XX XX, 2024; revised XX XX, XXXX.}}

\markboth{JOURNAL OF IEEE TRANSACTIONS ON DATA ENGINEERING, VOL. XX, NO. XX, XXXX}%
{Zhangchi Zhu \MakeLowercase{\textit{et al.}}: Preference-Consistency Distillation}


\maketitle

\begin{abstract}
Feature-based knowledge distillation has been applied to compress modern recommendation models, usually with projectors that align student (small) recommendation models' dimensions with teacher dimensions.
However, existing studies have only focused on making the projected features (i.e., student features after projectors) similar to teacher features, overlooking investigating whether the user preference can be transferred to student features (i.e., student features before projectors) in this manner.
In this paper, we find that due to the lack of restrictions on projectors, the process of transferring user preferences will likely be interfered with. We refer to this phenomenon as preference inconsistency. It greatly wastes the power of feature-based knowledge distillation. To mitigate preference inconsistency, we propose \ourmethod, which consists of two regularization terms for projectors. We also propose a hybrid method that combines the two regularization terms. We focus on items with high preference scores and significantly mitigate preference inconsistency, improving the performance of feature-based knowledge distillation. Extensive experiments on three public datasets and three backbones demonstrate the effectiveness of \ourmethod. The code of our method is provided in \url{https://github.com/woriazzc/KDs}.
\end{abstract}


\section{Introduction}
\IEEEPARstart{W}{ith} the rapid growth in the variety of items and the gradual diversification of user preferences, recommender systems play an increasingly important role. It has been found that small-capacity recommendation models are no longer applicable in today's big data era~\cite{ohsaka2023curse,liang2018variational}. To improve the recommendation performance of the model, these works propose to increase the embedding dimension of models.  However, due to the huge share of embeddings~\cite{ohsaka2023curse,zhao2023embedding} in the overall model parameters, increasing the embedding dimensionality greatly expands the number of parameters, which in turn incurs the high storage cost and inference latency, leading to longer waiting times and lower user satisfaction.

To improve the inference efficiency of recommendation models~\cite{kang2020rrd,chen2023unbiased,kang2021topology,kang2022personalized} without sacrificing their recommendation accuracy, knowledge distillation (KD) for recommender systems~ has attracted the attention of many researchers. KD is a model-agnostic approach for model compression~\cite{hinton2015distilling,gou2021knowledge}. It aims to transfer knowledge from the pre-trained large teacher to the small student to accelerate training and improve performance. Once training is complete, only the small student is used for inference, thus greatly improving inference efficiency. In knowledge distillation for recommendation, the common process is first to train a large teacher model using the user-item interactions, then train a small student model using the user-item interactions as well as the features in the intermediate layer~\cite{kang2020rrd,kang2021topology,kang2022personalized} and the predictions in the output layer~\cite{lee2019collaborative,kang2020rrd,kweon2021bidirectional,chen2023unbiased} provided by the teacher model. After training, the student trained with KD performs similarly to the teacher and obtains a much lower inference latency due to its small representation dimensionality. Most KD methods for recommendation~\cite{kweon2021bidirectional,tang2018ranking,chen2023unbiased,lee2019collaborative} are response-based methods that force the student model to learn the teacher's logits to improve student's performance. Recently, a growing body of research~\cite{romero2014fitnets,zagoruyko2016paying,passalis2018learning} suggests that the intermediate features of teacher models are rich in knowledge and can be used for knowledge distillation. These feature-based KD methods propose first utilizing a projector to align the dimensions of the student's and the teacher's features, then using the MSE loss to distill the features. Inspired by this, some feature-based knowledge distillation methods for recommender systems~\cite{kang2020rrd,kang2022personalized} are proposed. The state-of-the-art method, Distillation Experts (DE)~\cite{kang2020rrd}, first conducts a clustering on the features, then uses multiple projectors to distill the knowledge of each cluster separately. Recently, PHR~\cite{kang2022personalized} enables a personalized distillation for each user/item personalization and can be viewed as a generalization of DE.

\begin{figure}[!t]
\centering
  \includegraphics[width=\linewidth]{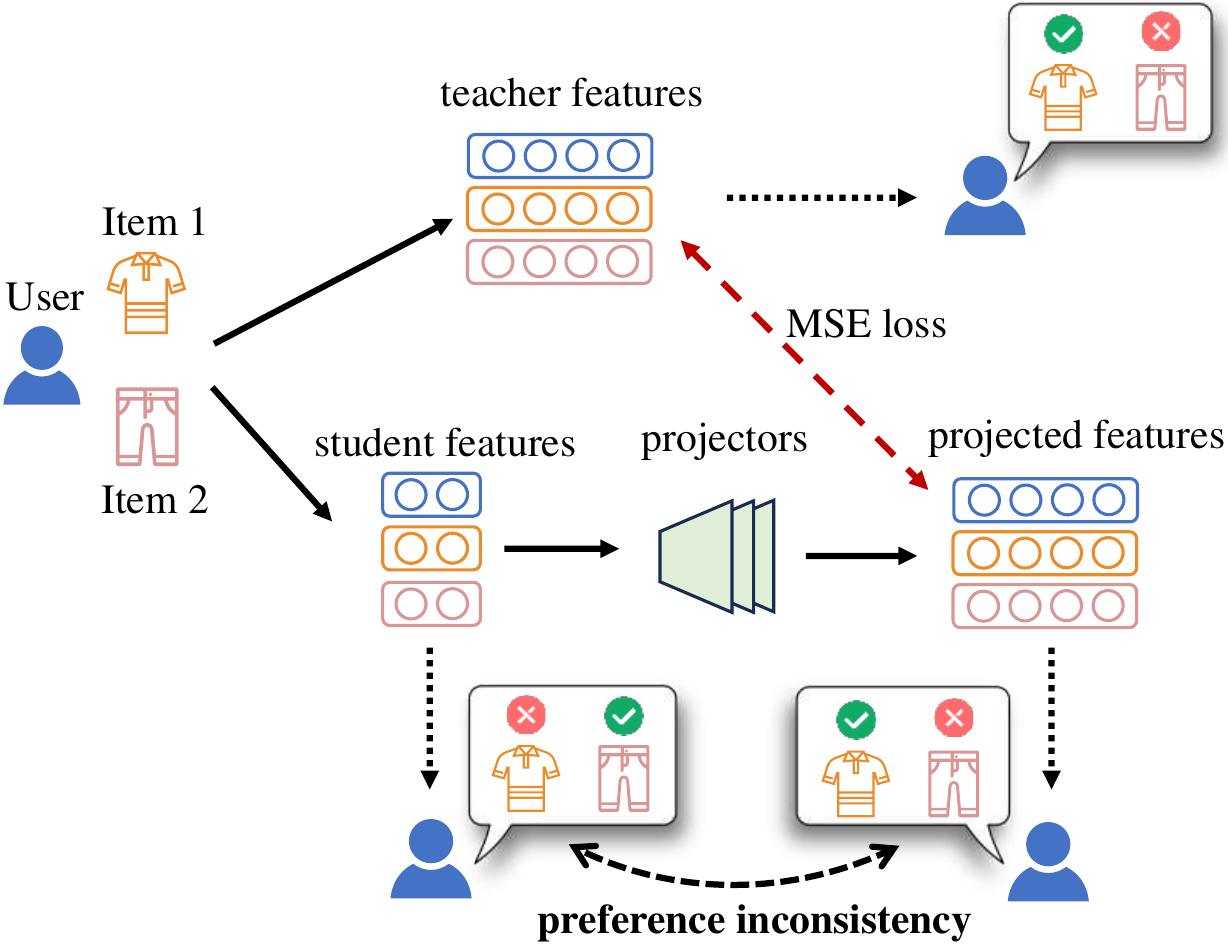}
  \caption{Illustration of feature-based knowledge distillation and preference inconsistency. Although the projected features derive the correct user preference (i.e., item 1 is preferred over item 2), the student does not benefit from it and instead derives the opposite user preference (i.e., item 2 is preferred over item 1).}
  \label{fig:intro1}
\end{figure}

Despite the effectiveness of existing feature-based knowledge distillation methods, they ignore a huge difference between the field of recommender systems and other fields, namely user preference. In recommender systems~\cite{rendle2012bpr,mao2021simplex,he2020lightgcn}, a recommendation model computes the user's preference score for each candidate item using user/item features, then ranks the items based on the scores. Thus, features are critical and should be conducive to getting the right rankings or user preferences, as they are called in this paper. Therefore, feature-based knowledge distillation should utilize the teacher's features so that the student's features can derive the correct user preferences. However, this may not be possible due to the introduction of projectors in feature-based knowledge distillation. As shown in Figure~\ref{fig:intro1}, existing feature-based methods~\cite{kang2020rrd,kang2022personalized} make the projected features (i.e., student's features after projectors) close to the teacher's features through the MSE loss to enable the derivation of correct user preferences (e.g., item 1 is preferred over item 2) using projected features. However, they assume that as long as projected features exhibit correct user preferences, the student features (i.e., those before projectors) have also learned correct user preferences. We argue that this is likely not the case due to the lack of restrictions on projectors.
This paper refers to this difference in rankings before and after projectors as \textbf{\textit{preference inconsistency}}. Preference inconsistency can be considered cheating by the student model via the projectors. Specifically, existing feature-based methods would assume that the student model has fully learned the teacher's knowledge about user preference (i.e., item 1 is preferred over item 2) because the projected features exhibit the same user preferences as the teacher. However, the fact is that students do not learn this knowledge (it demonstrates that item 2 is preferred over item 1). This is all because the student model camouflages its mastery through projectors; in other words, the student model cheats.

In our work, we empirically verify the existence of preference inconsistency and find that it is very prevalent even among items with high preference scores (see Figure~\ref{fig:method1} for details). Combined with the fact that high preference-scored items are essential in knowledge distillation for recommender systems as shown by existing works~\cite{kang2020rrd,kweon2021bidirectional}, we believe that preference inconsistency leads to a significant waste of the ability of feature-based knowledge distillation for recommender systems. To alleviate the preference inconsistency, we propose Preference-Consistent Knowledge Distillation (\ourmethod), which consists of two regularization terms for the projectors. We design these regularization terms inspired by the pair and list-wise loss in recommender systems. They have different recommendation accuracies (as shown in Table~\ref{tab:all}) and training costs (as shown in Table~\ref{tab:train_time} and Table~\ref{tab:train_gpu}), thus can be used in scenarios with different training cost constraints. Through extensive experiments, we show that these regularization terms can significantly reduce preference inconsistency and thus improve the performance of recommender systems.

To sum up, the key contributions of our work are as follows:
\begin{itemize}
    \item We point out the widespread problem of preference inconsistency in feature-based knowledge distillation for recommender systems. Moreover, we empirically demonstrate that it is severe even for items with high preference scores that play a key role in knowledge distillation.
    \item To mitigate preference inconsistency, we propose two regularization terms for projectors, namely pair-wise \ourmethod and list-wise \ourmethod. They can be used in scenarios with different training cost constraints. We also propose a hybrid method that combines the two regularization terms.
    \item We conduct extensive experiments to demonstrate the superiority of our method. Our method enables the student to perform comparably to the teacher with only one-tenth of the model size and one-third of the inference latency.
\end{itemize}

\section{Related Work}

\subsection{Knowledge Distillation}

Knowledge distillation (KD) is a model compression method that distills the knowledge from a well-trained teacher model to a small student model. Existing KD methods mainly fall into three categories~\cite{gou2021knowledge} based on their knowledge type: response-based methods, feature-based methods, and relation-based methods. \textit{Response-based} methods~\cite{hinton2015distilling,zhao2022decoupled,huang2022knowledge} take the final prediction of the teacher as soft targets, and the student is trained to mimic the soft targets by minimizing the divergence loss. \textit{Feature-based} methods take advantage of the teacher model's ability to learn multiple levels of feature representation with increasing abstraction. They propose to enrich auxiliary signals by matching the learned features of the student and the teacher. These methods leverage various types of knowledge, such as feature activations~\cite{romero2014fitnets}, attention maps~\cite{zagoruyko2016paying}, probability distribution in feature space~\cite{passalis2018learning}. \textit{Relation-based} methods~\cite{park2019relational,chen2020learning,passalis2020probabilistic} further explore the relationships between different layers or data samples. A representative example is RKD~\cite{park2019relational}, which calculates the similarity between each pair of samples to obtain a similarity matrix and then minimizes the difference between the teacher's and the student's similarity matrices.

\subsection{Knowledge Distillation in Recommender System}

KD has been introduced into recommender systems to reduce the representation dimensionality and inference latency. In recommender systems, knowledge distillation methods fall into the same three categories, i.e., response-based, feature-based, and relation-based.

\textit{Response-based} methods focus on the predictions of the teacher. For example, CD~\cite{lee2019collaborative} sample unobserved items from a distribution associated with their rankings predicted by the student: items with higher rankings are more likely to be sampled. RRD~\cite{kang2020rrd} adopts a list-wise loss to maximize the likelihood of the teacher's recommendation list. IR-RRD~\cite{kang2021item} further proposes to use item-side ranking loss to supplement user-side ranking information. DCD~\cite{lee2021dual} samples items and users from distributions associated with teacher-student prediction discrepancy to provide the student with dynamic knowledge. UnKD~\cite{chen2023unbiased} partitions items into multiple groups based on popularity and samples positive-negative item pairs within each group. HetComp~\cite{kang2023distillation} focuses on distillation ensembles of heterogeneous teachers and constructs a sequence of easy-to-difficult knowledge to narrow the discrepancy between the student model and heterogeneous teacher models. ALDI~\cite{huang2023aligning} exploits knowledge distillation in cold-start item recommendation to mitigate over-recommendation of warm or cold items. It treats warm items as teachers and cold items as students and designs several losses to align their ratings. Recently, DLLM2Rec~\cite{cui2024distillation} investigates knowledge distillation from Large Language Models (LLMs) to conventional sequential recommendation models. It is built based on RD~\cite{tang2018ranking} and adds a variety of methods for weighting instances.

\textit{Feature-based} methods focus on the intermediate representations of the teacher. For example, DE~\cite{kang2020rrd} proposes an expert module comprising $K$ projectors and a selection network. The selection network assigns each item or user to one of the projectors according to the teacher's representation of each. Then, the MSE loss is leveraged to align the representation of the teacher with the student representation transformed by the selected projector. PHR~\cite{kang2022personalized} employs a personalization network that enables a personalized distillation for each user/item representation, which can be viewed as a generalization of DE. DLLM2Rec~\cite{cui2024distillation} proposes to introduce a learnable offset term for each item to compensate for the huge gap in semantic space between LLM and traditional recommendation models.

\textit{Relation-based} methods focus on the relationships between different entities (i.e., users and items). PRecQ~\cite{shi2023quantize} trains a quantized small recommendation model by learning from item similarity relationships of a teacher recommendation model.
HTD~\cite{kang2021topology} observes that the student is too small to learn the whole relationship. As a result, the vanilla relation-based distillation approach is not always effective and even degrades the student’s performance. Therefore, it proposes to distill the sample relation hierarchically to alleviate the large capacity gap between the student and the teacher.

Our work belongs to feature-based methods. Although existing feature-based methods have shown good performance, they are often inferior to response-based methods. Existing feature-based methods only focus on designing a more personalized projector, ignoring the consistency of user preference before and after the projector, which leads to a large amount of wasted capacity of the projector. Our work is orthogonal to existing works. We point out the widespread existence of preference inconsistency and propose solutions.

\section{Preliminary}

\subsection{Top-N Recommendation}
In this work, we focus on the top-$N$ recommendation with implicit feedback. 
Specifically, let $\mathcal{U}$ and $\mathcal{I}$ denote the user set and item set, respectively. Then $|\mathcal{U}|$ and $|\mathcal{I}|$ are taken as the number of users and items, respectively. The historical implicit feedback can be formulated as a set of observed user-item interactions $\mathcal{R}=\{(u, i)|u$ interacted with $i\}$. A recommendation model aims to score the items not interacted with by the user and recommend $N$ items with the largest scores. 
Most recommendation models use an encoder network to map each user or item into a low-dimensional feature. For example, the encoder in matrix factorization models~\cite{koren2009matrix} is simply an embedding table, which takes the user ID or item ID as input and retrieves the corresponding embedding. The encoder in graph-based models~\cite{wang2019neural,he2020lightgcn} further enhances it with the neighbors' information. 
In this work, we use the bold symbols $\mathbf{u}$ and $\mathbf{i}$ to denote the feature of user $u$ and item $i$, respectively. 
Then, the predicted preference score is given by the matching function of the user and item features, such as the inner product. In this work, the matching function is denoted as $score(\mathbf{u},\mathbf{i})$, where $\mathbf{u}$ and $\mathbf{i}$ stand for the feature of the user and the item, respectively.
As for the learning objective, most works adopt the BPR loss~\cite{rendle2012bpr} to make the predicted scores of interacted items higher than randomly sampled negative items, which is given by
\begin{align}
    \mathcal{L}_{base}=\sum_{(u,i^+)\in \mathcal{R}, i^-}-\log \sigma\left(score(\mathbf{u},\mathbf{i}^+)-score(\mathbf{u},\mathbf{i}^-)\right)\,,\label{eq:Lbase}
\end{align}
where $\sigma(\cdot)$ is the sigmoid function, $i^+$ denotes an item interacted by user $u$, and $i^-$ is a randomly sampled negative item that $u$ has not interacted with.

\subsection{Feature-based Knowledge Distillation}
Let $d^s$ and $d^t$ denote the dimensionalities of student and teacher features, respectively. 
In this work, we use $\mathbf{u}^s\in \mathbb{R}^{d^s}$ and $\mathbf{i}^s\in\mathbb{R}^{d^s}$ to denote the student's features for user $u$ and item $i$, respectively. Similarly, the teacher's features for $u$ and $i$ are given by $\mathbf{u}^t\in\mathbb{R}^{d^t}$ and $\mathbf{i}^t\in\mathbb{R}^{d^t}$, respectively.

In feature-based knowledge distillation, the projector is first utilized to align the teacher and student models' dimensions, whereas when introduced into recommender systems~\cite{kang2020rrd,kang2022personalized}. Formally, consider the user $u$ and item $i$. Given the student's features $\mathbf{u}^s$ and $\mathbf{i}^s$, the projected features are defined as follows:
\begin{align*}
    \tilde{\mathbf{u}}^s:=E_U(\mathbf{u}^s),\\
    \tilde{\mathbf{i}}^s:=E_I(\mathbf{i}^s),
\end{align*}
where $E_U(\cdot)$ and $E_I(\cdot)$ are the projectors for user features and item features, respectively. Note that projectors can be a linear transformation~\cite{romero2014fitnets}, an MLP~\cite{kang2020rrd}, or an ensemble of MLPs~\cite{chen2022improved}.
Then, the student model is trained to minimize the MSE loss to make the student model mimic the teacher's features. And the feature-based knowledge distillation loss is given by
\begin{align}
\mathcal{L}_{FD}=\sum_{u\in\mathcal{U}}\|\tilde{\mathbf{u}}^s-\mathbf{u}^t\|_2+\sum_{i\in\mathcal{I}}\|\tilde{\mathbf{i}}^s-\mathbf{i}^t\|_2.\label{eq:loss_FD}
\end{align}

In this paper, we construct our method on top of DE~\cite{kang2020rrd} to compare it fairly. Specifically, we adopt $\mathcal{L}_{DE}$ proposed in DE instead of $\mathcal{L}_{FD}$. DE proposes to cluster users/items using the teacher's features first and then use different projectors for each cluster, which is orthogonal to our method. For specific details of DE, please refer to the original paper.

\section{Methodology}

\begin{figure*}[htbp]
\centering
  \includegraphics[width=\linewidth]{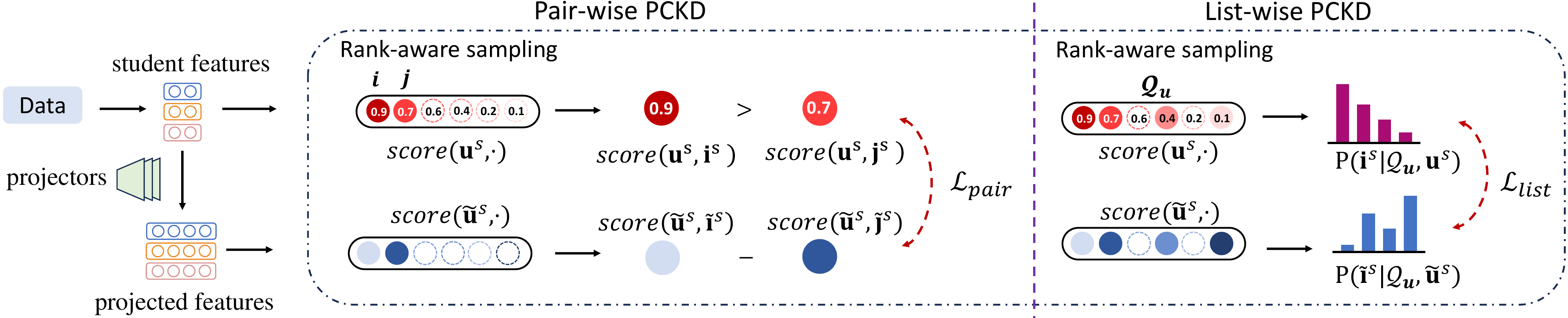}
  \caption{The process of \ourmethod. It consists of two regularization terms: pair-wise \ourmethod and list-wise \ourmethod. To compute them, we first conduct rank-aware sampling, followed by pair-wise or list-wise losses computed using the sampled items.}
  \label{fig:method}
\end{figure*}

In this section, we first analyze the preference inconsistency in Section~\ref{sec:anal}. Then, we propose \ourmethod to alleviate preference inconsistency in Section~\ref{sec:pckd}.
The illustration of \ourmethod is provided in Figure~\ref{fig:method}. \ourmethod consists of two regularization terms: pair-wise \ourmethod and list-wise \ourmethod. To compute them, we first conduct rank-aware sampling, followed by pair-wise or list-wise losses computed using the sampled items. In Section~\ref{sec:comb}, we further propose a hybrid method that combines the two regularization terms. Finally, in Section~\ref{sec:sum}, we summarize the training object.

\subsection{Preference Inconsistency in Feature-based Distillation}\label{sec:anal}

Given a recommendation model, the predicted preference scores are completely determined by its features for users and items. In other words, the features contain knowledge of user preferences, which is critical to the top-$N$ recommendation. However, we find that the knowledge contained in the features is highly likely to be corrupted by the projectors. In this section, we first formally define the preference inconsistency induced by projectors. Then, we empirically verify its widespread existence.

To define the preference inconsistency, we first provide the formal definition of preference, which has been implicitly considered in many existing works~\cite {rendle2012bpr,ren2021unbiased,rendle2014improving}.
\begin{definition}[Preference]\label{def:preference}
    In this work, we define the preference of each user in a pair-wise manner and compute it using the features. Specifically, given the feature of user $u$, item $i$, and item $j$, $u$'s preference for item $i$ and $j$ is defined as follows:
    \begin{align*}
        Pref(\mathbf{u},\mathbf{i},\mathbf{j}):=sign\left(score\left(\mathbf{u},\mathbf{i}\right)-score\left(\mathbf{u},\mathbf{j}\right)\right),
    \end{align*}
    where $sign(\cdot)$ is the sign function, which is given by
    \begin{align*}
        sign(x):=\begin{cases}
            -1 & \text{if $x<0$,}\\
            1 & \text{otherwise.}
        \end{cases}
    \end{align*}
\end{definition}
According to the above definition, the range of the values of $Pref(\mathbf{u},\mathbf{i},\mathbf{j})$ is $\{-1,1\}$. If $Pref(\mathbf{u},\mathbf{i},\mathbf{j})=1$, it means that the features tell us that user $u$ likes item $i$ more than item $j$, otherwise the opposite is true.

Given the definition of preference, we now formally define preference inconsistency.

\begin{definition}[Preference Inconsistency]
    For user $u$, the preference inconsistency associated with $u$ is given by
    \small{\begin{align*}
        C_{u}=\frac{1}{|\mathcal{I}|(|\mathcal{I}|-1)}\sum_{\substack{i\neq j,\\i,j\in\mathcal{I}}}\delta\left(Pref\left(\mathbf{u}^s,\mathbf{i}^s,\mathbf{j}^s\right)\neq Pref\left(\tilde{\mathbf{u}}^s, \tilde{\mathbf{i}}^s,\tilde{\mathbf{j}}^s\right)\right)
    \end{align*}}
    where 
    \begin{align*}
        \delta(x)=\begin{cases}
            1 & \text{if $x$ is True,}\\
            0 & \text{otherwise.}
        \end{cases}
    \end{align*}
    
    Since the direct computation of preference inconsistency requires traversing all $i\neq j$, its required time complexity is too high. Therefore, we approximate the computation by sampling.

    Furthermore, the preference inconsistency over the entire user set is given by
    \begin{align*}
        C=\frac{1}{|\mathcal{U}|}\sum_{u\in\mathcal{U}}C_u.
    \end{align*}
\end{definition}

Preference inconsistency is caused by projectors in feature-based knowledge distillation and can be considered cheating by the student model. For example, consider user $u$, item $i$, and item $j$. Suppose we have minimized the feature-based knowledge distillation loss in Eq.~(\ref{eq:loss_FD}) and make the projected features exactly equal to the teacher's features, i.e., $\tilde{\mathbf{u}}^s=\mathbf{u}^t$, $\tilde{\mathbf{i}}^s= \mathbf{i}^t$, and $\tilde{\mathbf{j}}^s=\mathbf{j}^t$. At this point, the knowledge about user preference contained in the teacher's features has been fully transferred to the projected features, i.e., $Pref(\tilde{\mathbf{u}}^s,\tilde{\mathbf{i}}^s,\tilde{\mathbf{j}}^s)=Pref(\mathbf{u}^t,\mathbf{i}^t,\mathbf{j}^t)$. Existing feature-based methods assume by default that projected features faithfully reflect the student model's knowledge about user preferences, i.e., $Pref(\mathbf{u}^s,\mathbf{i}^s,\mathbf{j}^s)=Pref(\tilde{\mathbf{u}}^s,\tilde{\mathbf{i}}^s,\tilde{\mathbf{j}}^s)$. Therefore, they assume that the student model has also fully learned the teacher's knowledge about user preferences. However, due to preference inconsistency (i.e., $Pref(\mathbf{u}^s,\mathbf{i}^s,\mathbf{j}^s)\neq Pref(\tilde{\mathbf{u}}^s,\tilde{\mathbf{i}}^s,\tilde{\mathbf{j}}^s)$), this is not the case. Thus, we can consider preference inconsistency as cheating by the student model via projectors. In other words, the student model uses erroneous projectors to make the projected features appear to match the teacher features rather than learning the knowledge about user preferences contained in the teacher features. From this perspective, eliminating preference inconsistency eliminates projectors that do not reflect the true state of the student model to prevent the student model from cheating through them.

\begin{figure}[!t]
\centering
  \includegraphics[width=\linewidth]{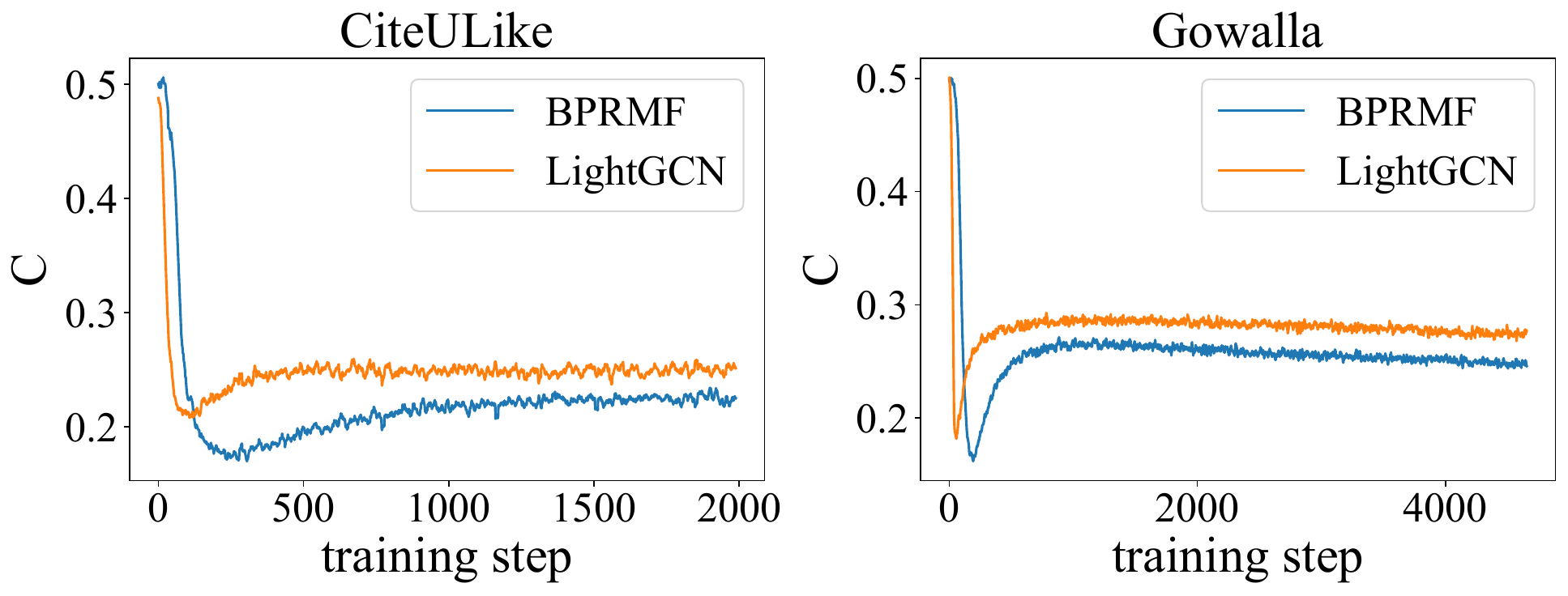}
  \caption{The dynamics of preference inconsistency as training proceeds. Experiments are conducted on CiteULike (left) and Gowalla (right).}
  \label{fig:method1}
\end{figure}

To verify the existence of preference inconsistency, we conduct experiments on CiteULike and Gowalla with BPRMF~\cite{rendle2012bpr} and LightGCN~\cite{he2020lightgcn} as the backbone. We report the preference inconsistency over the entire user set as training proceeds in Figure~\ref{fig:method1}. The results show that preference inconsistency goes through three stages as the training proceeds: (1) Preference inconsistency is very severe at the beginning of training, and users' preferences for almost half of the items flip before and after projectors. (2) Consequently, preference inconsistency decreases rapidly, and at the minimum, projectors interfere with about 20\% of user preferences on average. (3) Preference inconsistency begins to rise and eventually converges. When the training process is over, about 25\% of the user preferences are still interfered with by projectors. Based on these results, we find that preference inconsistency is always present throughout the training process and involves at least 20\% user preferences, and the model cannot reduce preference inconsistency by itself alone.

Since interactions in recommender systems usually follow a long-tailed distribution, i.e., the total number of items is large, but the number of items that are interacted with is small, it may be argued that preference inconsistency exists only for items that are virtually impossible to interact with, whereas those items with high preference scores are unaffected. Unfortunately, our next experiment refutes this. To investigate the association between preference inconsistency and the preference scores of the items, we first predict each user's preference scores for all items using the student model. Then, for each user (assuming user $u$), we sort all items according to their scores in descending order and divide items into five groups, denoted as $\{\mathcal{I}^u_1,\mathcal{I}^u_2,\cdots,\mathcal{I}^u_5\}$. After that, we define the group-wise preference inconsistency for $u$ as follows:
\begin{multline*}
    C_u^{m,n}=\frac{1}{N_{m,n}^u}\sum_{\substack{i\neq j,\\i\in\mathcal{I}_m^u,j\in\mathcal{I}_n^u}}\delta\left(Pref\left(\mathbf{u}^s,\mathbf{i}^s,\mathbf{j}^s\right)\right.\\
    \left.\neq Pref\left(\tilde{\mathbf{u}}^s, \tilde{\mathbf{i}}^s,\tilde{\mathbf{j}}^s\right)\right)\,,
\end{multline*}
where 
\begin{align*}
m\in\{1,2,\cdots,5\} \text{ and } n\in\{1,2,\cdots,5\},\\
    N_{m,n}^u=\begin{cases}
        |\mathcal{I}_m^u|(|\mathcal{I}_m^u|-1) & \text{if $m=n$},\\
        |\mathcal{I}_m^u|\cdot|\mathcal{I}_n^u| & \text{otherwise}.
    \end{cases}
\end{align*}

Finally, the group-wise preference inconsistency over the entire user set is defined as the average of $C_u^{m,n}$ for all users, which is given by
\begin{align*}
    C^{m,n}=\frac{1}{|\mathcal{U}|}\sum_{u\in\mathcal{U}}C_u^{m,n}, 
\end{align*}
where
\begin{align*}
    m\in\{1,2,\cdots,5\} \text{ and } n\in\{1,2,\cdots,5\}.
\end{align*}

Intuitively, $C^{m,n},\forall m\neq n$ measures the probability that a user's preference for two items is disturbed by projectors when their preference scores belong to different levels. On the other hand, $C^{m,n},\forall m=n$ measures the probability that a user's preference for two items is disturbed by projectors when their preference scores are similar (both high or both low).

\begin{figure}[htbp]
\centering
  \includegraphics[width=\linewidth]{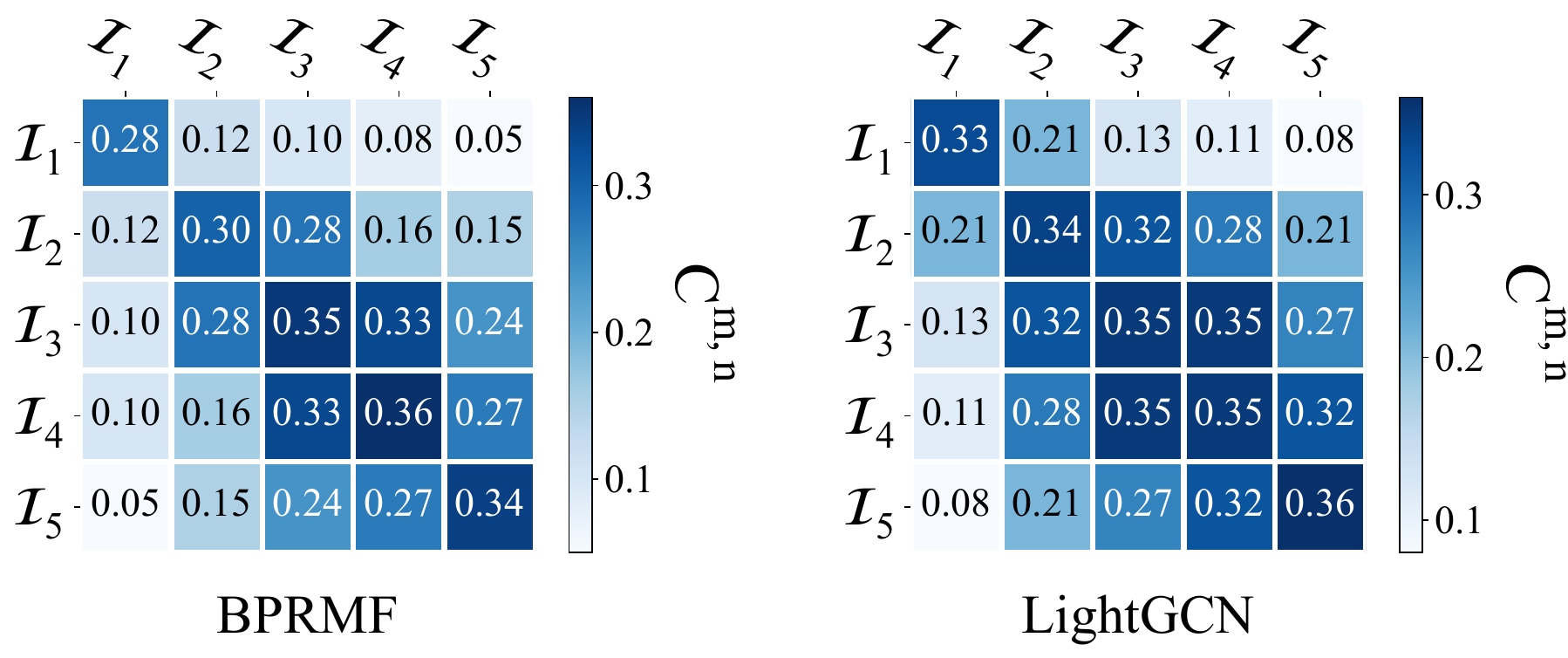}
  \caption{Group-wise preference inconsistency of DE.}
  \label{fig:method2}
\end{figure}

Figure~\ref{fig:method2} shows the group-wise preference inconsistency after convergence. From the results, we have two main findings:
\begin{itemize}[leftmargin=*]
\item Preference inconsistency is most severe when the preference scores for the two items are similar, i.e., $m=n$. Preference inconsistency eases as the gap in preference scores for the items increases. This is intuitive because small perturbations can flip user preferences when the preference scores of items are similar.
\item In the case of two items with similar preference scores, i.e., $m=n$, the lower the preference score, the more the items are affected by preference inconsistency. Nonetheless, items with the highest preference scores are still significantly affected. Specifically, $C^{1,1}$ is approximately equal to $0.3$, meaning that even if only those items most likely to be interacted with are considered, 30\% of user preferences are still disturbed.
\end{itemize}

In sum, preference inconsistency is widespread even among items with high preference scores. Note that, as has been shown in existing works~\cite{kang2020rrd,lee2019collaborative}, the influence of those items with high preference scores is significant in the knowledge distillation process. Therefore, we believe that reducing preference inconsistency in these items has significant implications for knowledge distillation.

\subsection{Preference-Consistent Knowledge Distillation}\label{sec:pckd}

According to the analysis in the previous section, preference inconsistency is widespread, even among items with high preference scores that play a key role in knowledge distillation. Therefore, in this section, we propose two regularization terms to reduce the preference inconsistency: pair-wise \ourmethod and list-wise \ourmethod. They have different recommendation accuracy and different training costs. Thus, they can be used in scenarios with different training cost constraints. Moreover, we perform ranking-aware sampling before computing the regularization terms to mitigate preference inconsistency in high-scored items.

\noindent\textbf{Pair-wise \ourmethod.}
As we analyzed in the previous section, eliminating preference inconsistency actually weeds out projectors that do not truly reflect the student's state. In other words, we should design regularization terms by targeting the student model's knowledge of user preferences and ensuring that the knowledge contained in the projected features is consistent with the target. Inspired by BPR loss~\cite{rendle2012bpr}, a well-known pair-wise loss for ranking, we propose $\mathcal{L}_{PCKD-P}$, which is given by


\begin{multline}\label{eq:loss_pair}
    \mathcal{L}_{PCKD-P}=-\sum_{u\in\mathcal{U}}\log\sigma\left(Pref\left(\mathbf{u}^s,\mathbf{i}^s,\mathbf{j}^s\right)\right.\\
    \left.\cdot\left(score\left(\tilde{\mathbf{u}}^s,\tilde{\mathbf{i}}^s\right)-score\left(\tilde{\mathbf{u}}^s,\tilde{\mathbf{j}}^s\right)\right)\right)\,,
\end{multline}
where $\sigma(\cdot)$ is the sigmoid function, and $Pref\left(\mathbf{u}^s,\mathbf{i}^s,\mathbf{j}^s\right)$ is the preference computed using the student's features, following the definition in Definition~\ref{def:preference}, which is computed as follows:
\begin{align*}
    Pref(\mathbf{u}^s,\mathbf{i}^s,\mathbf{j}^s)=sign\left(score\left(\mathbf{u}^s,\mathbf{i}^s\right)-score\left(\mathbf{u}^s,\mathbf{j}^s\right)\right).
\end{align*}

To compute $\mathcal{L}_{PCKD-P}$, we must sample two items for each user, i.e., $i$ and $j$. One baseline method can be random sampling, which equalizes preference inconsistencies across all items. On the other hand, existing work~\cite{kang2020rrd,lee2019collaborative} suggests that high-scoring items play a more critical role in knowledge distillation than other items. Therefore, we propose to sample $i$ and $j$ from the ranking-aware distributions similar to those proposed in CD~\cite{lee2019collaborative} so that higher-scoring items are more likely to be sampled. Specifically, we first sort all items according to the student's predicted preference scores. Then, the probability of the $k$-th item being sampled is given by
\begin{align}\label{eq:sample}
    p_u(k)\propto e^{-k/T},
\end{align}
where $T$ is the hyperparameter used to control the slope of the exponential function.

Note that the $Pref(\mathbf{u}^s,\mathbf{i}^s,\mathbf{j}^s)$ takes the value of $1$ or $-1$. When $Pref(\mathbf{u}^s,\mathbf{i}^s,\mathbf{j}^s)=1$, it indicates that the preference score of $u$ for $i$ is predicted to be greater than the score of $u$ for $j$ based on the student's features. At this point, $\mathcal{L}_{PCKD-P}$ makes $score\left(\tilde{\mathbf{u}}^s,\tilde{\mathbf{i}}^s\right)-score\left(\tilde{\mathbf{u}}^s,\tilde{\mathbf{j}}^s\right)$ larger, which in turn drives $i$'s score computed from projected features to be larger than $j$'s score, thus consistent with the prediction of the student features. When $Pref(\mathbf{u}^s,\mathbf{i}^s,\mathbf{j}^s)=-1$, it can be analyzed similarly to find that $\mathcal{L}_{PCKD-P}$ aligns user preferences predicted based on student's features with those predicted based on the projected features.

\noindent\textbf{List-wise \ourmethod.}
In addition to pair-wise loss, list-wise loss is also widely used in recommender systems due to its computational simplicity, fast convergence, and ability to address the vanishing gradients while considering class relationships. Inspired by the cross-entropy (CE) loss function~\cite{kang2020rrd}, we also provide a list-wise loss to align the preference distribution computed from the student's features and that computed from the projected features.
Before calculating the loss, we independently sample $Q$ items for each user, forming the set $\mathcal{Q}_u$. The sampling function is the same as in pair-wise \ourmethod, given in Eq.(\ref{eq:sample}).

Then, the list-wise regularization term is defined as follows
\begin{align}\label{eq:loss_list}
    \mathcal{L}_{PCKD-L} = -\sum_{u\in \mathcal{U}}\sum_{i\in \mathcal{Q}_u}\text{P}(\mathbf{i}^s|\mathcal{Q}_u, \mathbf{u}^s)\log \text{P}\left(\tilde{\mathbf{i}}^s|\mathcal{Q}_u, \tilde{\mathbf{u}}^s\right)\,,
\end{align}
where $\text{P}(\mathbf{i}^s|\mathcal{Q}_u, \mathbf{u}^s)$ and $\text{P}\left(\tilde{\mathbf{i}}^s|\mathcal{Q}_u, \tilde{\mathbf{u}}^s\right)$ denote the preference distribution computed from the student's features and the projected features, respectively. Formally, they are given by
\begin{align}
    \text{P}(\mathbf{i}^s|\mathcal{Q}_u, \mathbf{u}^s)&=\frac{\exp\left(score(\mathbf{u}^s,\mathbf{i}^s)\right)}{\sum_{j\in \mathcal{Q}_u}\exp\left(score\left(\mathbf{u}^s,\mathbf{j}^s\right)\right)}\,,\\
    \text{P}\left(\tilde{\mathbf{i}}^s|\mathcal{Q}_u, \tilde{\mathbf{u}}^s\right)&=\frac{\exp\left(score\left(\tilde{\mathbf{u}}^s,\tilde{\mathbf{i}}^s\right)\right)}{\sum_{j\in \mathcal{Q}_u}\exp\left(score\left(\tilde{\mathbf{u}}^s,\tilde{\mathbf{j}}^s\right)\right)}\,.
\end{align}

By targeting the preference distribution computed from the student's features and minimizing the CE loss, $\mathcal{L}_{PCKD-L}$ keeps the preference distribution the same before and after the projectors.

Notice that different numbers of items are involved in the calculation of $\mathcal{L}_{PCKD-P}$ and $\mathcal{L}_{PCKD-L}$. In addition, $\mathcal{L}_{PCKD-P}$ focuses on the relative order of the items‘ scores, while $\mathcal{L}_{PCKD-L}$ aims to make the distribution of the items’ scores unchanged, so the two losses play different roles. When choosing from these two losses, a trade-off must be made as list-wise loss usually performs better but requires higher computational costs. In the experiment section, we
 provide details of the training cost and the accuracy of the recommendation for these two regularization terms.

\subsection{Combining Pair-wise and List-wise PCKD}\label{sec:comb}

While we intended to design two different regularization terms to accommodate scenarios with different training budgets, readers may wonder if it is possible to combine these two regularization terms to further improve recommendation performance. In this section, we propose a hybrid \ourmethod, denoted as PCKD-H, to combine pair-wise and list-wise PCKD.

In computing PCKD-H, we propose to modify the sampling strategy in pair-wise loss (i.e., PCKD-P) to make pair-wise and list-wise PCKD responsible for items of different predictive scores. Since our method samples items that are typically highly ranked, the individual regularization terms proposed in the previous section will focus on items with similar rankings (all with high rankings) and ignore preference inconsistencies between items with large differences in rankings. Therefore, we propose to compensate for this by modifying the sampling strategy in pair-wise PCKD (i.e., PCKD-P) when combining pair-wise and list-wise PCKD. Specifically, 
\begin{itemize}
    \item For the list-wise loss (i.e., PCKD-L), we do not make any changes. It is responsible for improving preference inconsistency in top items.
    \item For the pair-wise loss (i.e., PCKD-P), in the original paper, the two items used to calculate it are sampled from the same distribution. We modify it to sample the two items from two distributions given by different hyperparameters $T$ respectively. Formally, the first item (denoted as $i$) is sampled from the distribution of $p_{u,i}(k)\sim e^{-k/T_1}$, while the second item (denoted as $j$) is sampled from $p_{u,j}(k)\sim e^{-k/T_2}$. By setting $T_1<T_2$, we expect item $i$ to be a top item, whereas item $j$ is more likely to be a non-top item. Therefore, we can ensure that there will be at least one top item while allowing for a greater difference in the predictive scores between sampled items.
\end{itemize}
Finally, we compute PCKD-H by combining PCKD-L with modified PCKD-P through a hyperparameter $\alpha$ and
\begin{align}\label{eq:loss_comb}
\mathcal{L}_{PCKD-H}=(1 - \alpha)\cdot\mathcal{L}_{PCKD-L}+\alpha\cdot \mathcal{L}_{PCKD-P}',
\end{align}
where $\mathcal{L}_{PCKD-P}'$ denotes the modified PCKD-P.

\subsection{Overall Training Objective}\label{sec:sum}

\ourmethod is jointly optimized with the base model's loss function and the feature-based distillation loss in an end-to-end manner as follows:
\begin{align}
    \mathcal{L}=\mathcal{L}_{base} + \lambda_{DE}\cdot \mathcal{L}_{DE}+\lambda_{\ourmethod}\cdot\mathcal{L}_{\ourmethod},\label{eq:all}
\end{align}
where $\mathcal{L}_{base}$ is the loss function of the base recommendation model in Eq.~(\ref{eq:Lbase}) and $\mathcal{L}_{DE}$ the feature-based distillation loss proposed in DE~\cite{kang2020rrd}. $\lambda_{DE}$ and $\lambda_{\ourmethod}$ are hyperparameters that control the effect of $\mathcal{L}_{DE}$ and \ourmethod, respectively. $\mathcal{L}_{\ourmethod}$ can be either the pair-wise loss $\mathcal{L}_{PCKD-P}$ in Eq.~(\ref{eq:loss_pair}), the list-wise loss $\mathcal{L}_{PCKD-L}$ in Eq.~(\ref{eq:loss_list}), or the combined loss $\mathcal{L}_{PCKD-H}$ in Eq.~(\ref{eq:loss_comb}).

The supplementary material exhibits the pseudo-code for the overall training process. Since updating the rankings given by the student every epoch is time-consuming, we conduct this step every $K$ epoch (lines 3-4 in the pseudo-code). In our experiments, we fix $K$ to $5$.

\section{Experiments}
In this section, we conduct experiments on three public datasets using three base models to validate the effectiveness of \ourmethod. The overall performance comparison is presented in Section~\ref{sec:result}. Section~\ref{sec:impact} analyses the impact of our method on preference inconsistency. The ablation study is provided in Section~\ref{sec:abl}. The analysis of training efficiency is presented in Section~\ref{sec:eff}. The analysis of inference efficiency is presented in Section 2 in the supplementary material. Section~\ref{sec:htd} investigates the preference inconsistency in HTD, a relation-based method. In Section~\ref{sec:response}, we report the impact of response-based methods on preference inconsistency. Section~\ref{sec:pckd-h} reports the performance of \ourmethod-H, the hybrid method. The hyperparameter analysis is provided in Section~\ref{sec:hyper}.

\subsection{Experimental Settings}

\begin{table}[htbp]
    \caption{Statistics of the preprocessed datasets.}
    \centering
    \begin{tabular}{ccccc}\toprule
         Dataset & \#Users & \#Items & \#Interactions & \#Sparsity\\
         \midrule
         \texttt{CiteULike} & 5,219 & 25,181 & 125,580 & 99.89\%\\
         \texttt{Gowalla} & 29,858 & 40,981 & 1,027,370 & 99.92\%\\
         \texttt{Yelp2018} & 41,801 & 26,512 & 1,022,604 & 99.91\%\\
         \bottomrule
    \end{tabular}
    \label{tab:dataset}
\end{table}

\begin{table*}[htbp]
    \caption{Dimensionalities of teachers and students.}
    \label{tab:dim}
    \centering
    \begin{tabular}{c|cccc|cccc|cccc} \toprule
        \multirow{2}{*}{Method} & \multicolumn{4}{c|}{CiteULike} & \multicolumn{4}{c|}{Gowalla} & \multicolumn{4}{c}{Yelp}\\
         & BPRMF & LightGCN & SimpleX & JGCF & BPRMF & LightGCN & SimpleX & JGCF & BPRMF & LightGCN & SimpleX & JGCF\\
        \midrule
        Teacher & 400 & 2000 & 400 & 1000 & 300 & 2000 & 1000 & 1000 & 300 & 1000 & 500 & 1000\\
        Student & 20 & 20 & 20 & 20 & 20 & 20 & 20 & 20 & 20 & 20 & 20 & 20\\
        \bottomrule
    \end{tabular}
\end{table*}

\noindent\textbf{Datasets.}
We conduct experiments on three public datasets, including \textbf{CiteULike}\footnote{\url{https://github.com/changun/CollMetric/tree/master/citeulike-t}}~\cite{wang2013collaborative,kang2022personalized,kang2021topology}, \textbf{Gowalla}\footnote{\url{http://dawenl.github.io/data/gowalla pro.zip}}~\cite{cho2011friendship,tang2018ranking,lee2019collaborative}, and \textbf{Yelp2018}\footnote{\url{https://github.com/hexiangnan/sigir16-eals}}~\cite{lee2019collaborative,kweon2021bidirectional}. We split training and test datasets following the previous method~\cite{xu2023stablegcn}. Specifically, we filter out users and items with less than 10 interactions and then split the rest chronologically into training, validation, and test sets in a ratio of 8:1:1. The statistics of the preprocessed datasets are summarized in Table~\ref{tab:dataset}.

\noindent\textbf{Evaluation Protocols.}
Per the custom, we adopt the full-ranking evaluation to achieve an unbiased evaluation. To evaluate the performance of top-$N$ recommendation, we employ Recall (Recall@$N$) and normalized discounted cumulative gain (NDCG@$N$) and report the results for $N \in \{10, 20\}$. We conduct five independent runs for each configuration and report the averaged results.

\noindent\textbf{Baselines.}
We compare our method with all three types of knowledge distillation methods:
\begin{itemize}[leftmargin=*]
\item For \textit{response-based} methods, we choose \textbf{RRD}~\cite{kang2020rrd} and \textbf{UnKD}\cite{chen2023unbiased}. They both use the teacher's logits to provide additional supervision to the student.
\item For \textit{relation-based} method, we select \textbf{HTD}~\cite{kang2021topology}. It uses the selector network to partition the samples and distill the sample relations hierarchically.
\item For \textit{feature-based} methods, we select \textbf{DE}~\cite{kang2020rrd} and \textbf{FitNet}~\cite{romero2014fitnets}. They belong to the same category as our approach, and all use the teacher model's features.
\end{itemize}

\noindent\textbf{Backbones.}
We refer to previous works~\cite{chen2023unbiased,kang2020rrd,kang2021topology} and choose BPRMF and LightGCN. We also add SimpleX as a new backbone.
\begin{itemize}[leftmargin=*]
\item \textbf{BPRMF}~\cite{rendle2012bpr} prompts the observed items to get higher preference scores than unobserved items through a pair-wise ranking loss function.
\item \textbf{LightGCN}~\cite{he2020lightgcn} is an efficient GCN for recommender systems that utilize neighborhood aggregation to enhance the features.
\item \textbf{SimpleX}~\cite{mao2021simplex} proposes cosine contrastive loss to maximize the similarity between positive pairs and minimize the similarity of negative pairs below a margin.
\item 
\textbf{JGCF}~\cite{guo2023manipulating} is a graph recommendation model that uses Jacobi polynomial bases to emphasize the low and high frequencies and controls the strength of the mid-frequencies using hyperparameters.

\end{itemize}

\noindent\textbf{Teacher/Student.}
Our work focuses on scenarios where teachers and students have the same backbone. For each backbone, we increase the model size until the recommendation performance is no longer improved and adopt the model with the best performance as the teacher model. We set the representation dimensionality of the student model to 20. The dimensionalities of the teachers and the students are summarized in Table~\ref{tab:dim}.

\noindent\textbf{Implementation Details.}
We implement all the methods with PyTorch and use Adam as the optimizer in all experiments. For our method, the weight decay is selected from \{1e-3, 1e-4, 1e-5, 0\}. The search space of the learning rate is \{1e-3, 1e-4\}. $Q$ is selected from \{5, 10\}, $T$ is fixed to $10$ as we find it has little influence on the performance, and $\lambda_{\ourmethod}$ is selected from \{1e-3, 5e-3, 1e-2\}. For a fair comparison with DE, we fix the hyperparameters of DE after tuning them to the optimum and then tune the hyperparameters in our method. We set the total number of training epochs as 1000. We also conduct early stopping according to the NDCG@20 on the validation set and stop training when the NDCG@20 doesn't increase for 30 consecutive epochs. All hyperparameters of the compared baselines are tuned in the same way as reported in their works.

\subsection{Performance Comparison}\label{sec:result}

\begin{table*}[!t]
    \caption{Recommendation performance. \ourmethod-P and \ourmethod-L denote the pair-wise \ourmethod and list-wise \ourmethod, respectively. The best results are in boldface, and the best baselines are underlined. \textit{Improv.b} denotes the relative improvement of \ourmethod over the best baseline. \textit{Improv.f} denotes the relative improvement of \ourmethod over the best feature-based distillation method. A paired t-test is performed over 5 independent runs to evaluate $p$-value. * and ** indicate $p\le 0.05$ and $p\le 0.005$, respectively.}
    \label{tab:all}
    \centering
    \resizebox{\linewidth}{!}{
    \begin{tabular}{cc|cccc|cccc|cccc} \toprule
        \multirow{2}{*}{Backbone} & \multirow{2}{*}{Method} & \multicolumn{4}{c|}{CiteULike} & \multicolumn{4}{c|}{Gowalla} & \multicolumn{4}{c}{Yelp}\\
         & & R@10 & N@10 & R@20 & N@20 & R@10 & N@10 & R@20 & N@20 & R@10 & N@10 & R@20 & N@20\\
        \midrule
        \multirow{11}{*}{BPRMF} & Teacher & 0.0283 & 0.0155 & 0.0442 & 0.0198 & 0.1088 & 0.0907 & 0.1544 & 0.1053 & 0.0394 & 0.0253 & 0.0660 & 0.0339\\
         & Student & 0.0177 & 0.0098 & 0.0284 & 0.0128 & 0.0946 & 0.0820 & 0.1329 & 0.0939 & 0.0348 & 0.0222 & 0.0586 & 0.0299\\
         \cline{2-14}
         & RRD & \underline{0.0235} & \underline{0.0135} & 0.0342 & \underline{0.0165} & 0.0977 & \underline{0.0861} & 0.1395 & \underline{0.0987} & 0.0362 &\underline{0.0235} & \underline{0.0625} & \underline{0.0319}\\
         & UnKD & 0.0232 & 0.0129 & \underline{0.0345} & 0.0160 & \underline{0.0981} & 0.0860 & \underline{0.1401} & 0.0979 & \underline{0.0366} & 0.0228 & 0.0601 & 0.0312\\
         & HTD & 0.0223 & 0.0127 & 0.0337 & 0.0152 & 0.0975 & 0.0852 & 0.1379 & 0.0970 & 0.0360 & 0.0231 & 0.0600 & 0.0309\\
         & DE & 0.0214 & 0.0120 & 0.0330 & 0.0147 & 0.0969 & 0.0847 & 0.1372 & 0.0971 & 0.0355 & 0.0227 & 0.0598 & 0.0305\\
         & FitNet & 0.0199 & 0.0108 & 0.0326 & 0.0140 & 0.0963 & 0.0841 & 0.1352 & 0.0967 & 0.0355 & 0.0226 & 0.0599 & 0.0301\\
         \cline{2-14}
         & \ourmethod-P & 0.0231 & 0.0133 & 0.0347 & 0.0160 & 0.0991 & 0.0859 & 0.1420 & 0.0983 & 0.0365 & 0.0235 & 0.0623 & 0.0317\\
         & \ourmethod-L & \textbf{0.0240} & \textbf{0.0144} & \textbf{0.0355} & \textbf{0.0172} & \textbf{0.1002} & \textbf{0.0877} & \textbf{0.1445} & \textbf{0.1006} & \textbf{0.0378} & \textbf{0.0247} & \textbf{0.0633} & \textbf{0.0329}\\
         \cline{2-14}
         & \textit{Improv.b} & 2.13\%** & 6.67\%** & 2.90\%** & 4.24\%** & 2.14\%* & 1.86\%* & 3.14\%** & 1.93\%* & 3.28\%** & 5.11\%** & 1.28\%* & 3.13\%**\\
         & \textit{Improv.f} & 12.15\%** & 20.00\%** & 7.58\%** & 17.01\%** & 3.41\%** & 3.54\%** & 5.32\%** & 3.60\%** & 6.48\%** & 8.81\%** & 5.85\%** & 7.87\%**\\
        \midrule
        \midrule
         \multirow{11}{*}{LightGCN} & Teacher & 0.0296 & 0.0160 & 0.0461 & 0.0205 & 0.1236 & 0.1035 & 0.1730 & 0.1190 & 0.0432 & 0.0276 & 0.0716 & 0.0367\\
         & Student & 0.0215 & 0.0113 & 0.0344 & 0.0148 & 0.1098 & 0.0928 & 0.1550 & 0.1069 & 0.0363 & 0.0235 & 0.0621 & 0.0308\\
         \cline{2-14}
         & RRD & 0.0231 & \underline{0.0125} & \underline{0.0359} & \underline{0.0158} & \underline{0.1142} & \underline{0.0969} & \underline{0.1627} & 0.1109 & \underline{0.0381} & \underline{0.0245} & \underline{0.0671} & \underline{0.0334}\\
         & UnKD & \underline{0.0233} & 0.0122 & 0.0353 & 0.0157 & 0.1138 & 0.0967 & 0.1622 & \underline{0.1112} & 0.0379 & 0.0242 & 0.0658 & 0.0329\\
         & HTD & 0.0225 & 0.0120 & 0.0352 & 0.0154 & 0.1139 & 0.0951 & 0.1589 & 0.1102 & 0.0375 & 0.0245 & 0.0653 & 0.0330\\
         & DE & 0.0220 & 0.0115 & 0.0349 & 0.0153 & 0.1111 & 0.0943 & 0.1578 & 0.1094 & 0.0369 & 0.0242 & 0.0641 & 0.0319\\
         & FitNet & 0.0219 & 0.0114 & 0.0347 & 0.0150 & 0.1103 & 0.0939 & 0.1571 & 0.1088 & 0.0368 & 0.0238 & 0.0633 & 0.0321\\
         \cline{2-14}
         & \ourmethod-P & 0.0227 & 0.0121 & 0.0360 & 0.0157 & 0.1141 & 0.0967 & 0.1637 & 0.1107 & 0.0376 & 0.0244 & 0.0660 & 0.0330\\
         & \ourmethod-L & \textbf{0.0238} & \textbf{0.0128} & \textbf{0.0371} & \textbf{0.0165} & \textbf{0.1165} & \textbf{0.0991} & \textbf{0.1663} & \textbf{0.1133} & \textbf{0.0394} & \textbf{0.0249} & \textbf{0.0688} & \textbf{0.0342}\\
         \cline{2-14}
         & \textit{Improv.b} & 2.15\%* & 2.40\%* & 3.24\%** & 4.43\%** & 2.01\%** & 2.27\%* & 2.21\%* & 1.89\%* & 3.41\%** & 1.63\%* & 2.53\%** & 2.40\%*\\
         & \textit{Improv.f} & 8.18\%** & 11.30\%** & 6.30\%** & 7.84\%** & 4.86\%** & 5.09\%** & 5.39\%** & 3.56\%** & 6.78\%** & 2.89\%** & 7.33\%** & 6.54\%**\\
         \midrule
         \midrule
         \multirow{11}{*}{SimpleX} & Teacher & 0.0343 & 0.0191 & 0.0508 & 0.0236 & 0.1184 & 0.0943 & 0.1750 & 0.1122 & 0.0469 & 0.0303 & 0.0778 & 0.0402\\
         & Student & 0.0290 & 0.0162 & 0.0426 & 0.0199 & 0.1046 & 0.0855 & 0.1524 & 0.1006 & 0.0382 & 0.0241 & 0.0658 & 0.0330\\
         \cline{2-14}
         & RRD & \underline{0.0323} & \underline{0.0175} & \underline{0.0466} & 0.0219 & \underline{0.1101} & \underline{0.0893} & \underline{0.1615} & 0.1059 & \underline{0.0422} & 0.0268 & \underline{0.0717} & 0.0361\\
         & UnKD & 0.0321 & 0.0175 & 0.0459 & \underline{0.0223} & 0.1081 & 0.0887 & 0.1607 & \underline{0.1061} & 0.0413 & \underline{0.0270} & 0.0713 & \underline{0.0362}\\
         & HTD & 0.0317 & 0.0172 & 0.0462 & 0.0215 & 0.1089 & 0.0883 & 0.1599 & 0.1057 & 0.0406 & 0.0258 & 0.0690 & 0.0353\\
         & DE & 0.0313 & 0.0165 & 0.0449 & 0.0203 & 0.1073 & 0.0869 & 0.1567 & 0.1042 & 0.0403 & 0.0253 & 0.0671 & 0.0347\\
         & FitNet & 0.0305 & 0.0167 & 0.0434 & 0.0205 & 0.1064 & 0.0862 & 0.1558 & 0.1031 & 0.0398 & 0.0249 & 0.0667 & 0.0341\\
         \cline{2-14}
         & \ourmethod-P & 0.0321 & 0.0176 & 0.0470 & 0.0220 & 0.1110 & 0.0899 & 0.1632 & 0.1064 & 0.0430 & 0.0266 & 0.0710 & 0.0358\\
         & \ourmethod-L & \textbf{0.0331} & \textbf{0.0180} & \textbf{0.0481} & \textbf{0.0229} & \textbf{0.1139} & \textbf{0.0919} & \textbf{0.1683} & \textbf{0.1088} & \textbf{0.0439} & \textbf{0.0279} & \textbf{0.0731} & \textbf{0.0372}\\
         \cline{2-14}
         & \textit{Improv.b} & 3.12\%** & 2.86\%** & 3.22\%** & 2.69\%** & 3.45\%** & 2.91\%** & 4.21\%** & 2.54\%** & 4.03\%** & 3.33\%** & 1.95\%* & 2.76\%*\\
         & \textit{Improv.f} & 5.75\%** & 9.09\%** & 7.13\%** & 11.71\%** & 6.15\%** & 5.75\%** & 7.40\%** & 4.14\%** & 8.93\%** & 10.28\%** & 8.94\%** & 7.20\%**\\
        \midrule
        \midrule
        \multirow{11}{*}{JGCF} & Teacher & 0.0325 & 0.0187 & 0.0492 & 0.0243 & 0.1283 & 0.1049 & 0.1803 & 0.1230 & 0.0455 & 0.0312 & 0.0783 & 0.0412\\
         & Student & 0.0277 & 0.0129 & 0.0392 & 0.0183 & 0.1108 & 0.0943 & 0.1592 & 0.1087 & 0.0379 & 0.0251 & 0.0670 & 0.0340\\
         \cline{2-14}
         & RRD & \underline{0.0301} & 0.0149 & 0.0437 & \underline{0.0206} & 0.1172 & \underline{0.0986} & 0.01671 & \underline{0.1134} & 0.0405 & \underline{0.0283} & \underline{0.0728} & \underline{0.0373}\\
         & UnKD & 0.0297 & \underline{0.0152} & \underline{0.0437} & 0.0201 & \underline{0.1185} & 0.0971 & \underline{0.1678} & 0.1130 & \underline{0.0406} & 0.0273 & 0.0711 & 0.0370\\
         & HTD & 0.0289 & 0.0145 & 0.0430 & 0.0198 & 0.1148 & 0.0969 & 0.1640 & 0.1118 & 0.0390 & 0.0270 & 0.0687 & 0.0368\\
         & DE & 0.0284 & 0.0144 & 0.0425 & 0.0195 & 0.1149 & 0.0957 & 0.1624 & 0.1114 & 0.0388 & 0.0269 & 0.0689 & 0.0360\\
         & FitNet & 0.0281 & 0.0139 & 0.0410 & 0.0195 & 0.1133 & 0.0952 & 0.1610 & 0.1109 & 0.0387 & 0.0265 & 0.0681 & 0.0353\\
         \cline{2-14}
         & PCKD-P & 0.0301 & 0.0150 & 0.0442 & 0.0206 & 0.1217 & 0.0997 & 0.1701 & 0.1160 & 0.0411 & 0.0285 & 0.0722 & 0.0378\\
         & PCKD-L & \textbf{0.0310} & \textbf{0.0156} & \textbf{0.0451} & \textbf{0.0213} & \textbf{0.1220} & \textbf{0.1008} & \textbf{0.1723} & \textbf{0.1165} & \textbf{0.0423} & \textbf{0.0288} & \textbf{0.0743} & \textbf{0.0385}\\
         \cline{2-14}
         & \textit{Improv.b} & 2.99\%** & 2.63\%** & 3.20\%** & 3.39\%** & 2.95\%** & 2.23\%** & 2.68\%** & 2.73\%** & 4.18\%** & 1.76\%* & 2.06\%** & 3.21\%**\\
         & \textit{Improv.f} & 9.15\%** & 8.33\%** & 6.11\%** & 9.23\%** & 6.17\%** & 5.32\%** & 6.09\%** & 4.50\%** & 9.02\%** & 7.06\%** & 7.83\%** & 6.94\%**\\
        \bottomrule
    \end{tabular}
    }
\end{table*}

Table~\ref{tab:all} shows the performance of \ourmethod and baselines over three public datasets and three backbones. From the results, we observe that 
our method significantly outperforms all baselines, especially feature-based and relation-based knowledge distillation methods. It even shows comparable performance to the teacher in many cases. Moreover, List-wise PCKD performs better than pair-wise PCKD. In more detail, we find that:

\begin{itemize}[leftmargin=*]
    \item List-wise \ourmethod outperforms all baselines by a large margin on all three datasets and three backbones and achieves remarkable improvements over the best baseline, especially on the CiteULike dataset. This demonstrates the effectiveness of \ourmethod.
    \item List-wise \ourmethod (\ourmethod-L) outperforms pair-wsie \ourmethod (\ourmethod-P). This is consistent with the observation in recent works~\cite{klenitskiy2023turning,xu2024fairly} that list-wise losses are better than pair-wise losses in recommender systems. Nevertheless, pair-wise PCKD still outperforms the best baseline in many cases.
    \item Both list-wise \ourmethod and pair-wise \ourmethod significantly outperform the feature-based distillation methods. Note that the only difference between our method and the feature-based method (DE) is eliminating preference inconsistency. Thus, this result verifies that reducing preference inconsistency can improve the performance of feature-based methods.
    \item Response-based knowledge distillation methods are always better than other baseline methods, including feature- and relation-based methods. Considering the excellent performance of feature-based knowledge distillation methods in CV and NLP~\cite{huang2022masked,liu2023norm,yang2022masked}, this reflects the wasted power of existing feature-based knowledge distillation methods for recommender systems.
    \item In many cases, \ourmethod shows comparable performance to the teacher, making it possible to achieve inference efficiency without sacrificing recommendation accuracy.
\end{itemize}

\subsection{Impact on Preference Inconsistency}\label{sec:impact}

\begin{figure}[!t]
  \centering
  \includegraphics[width=1.0\linewidth]{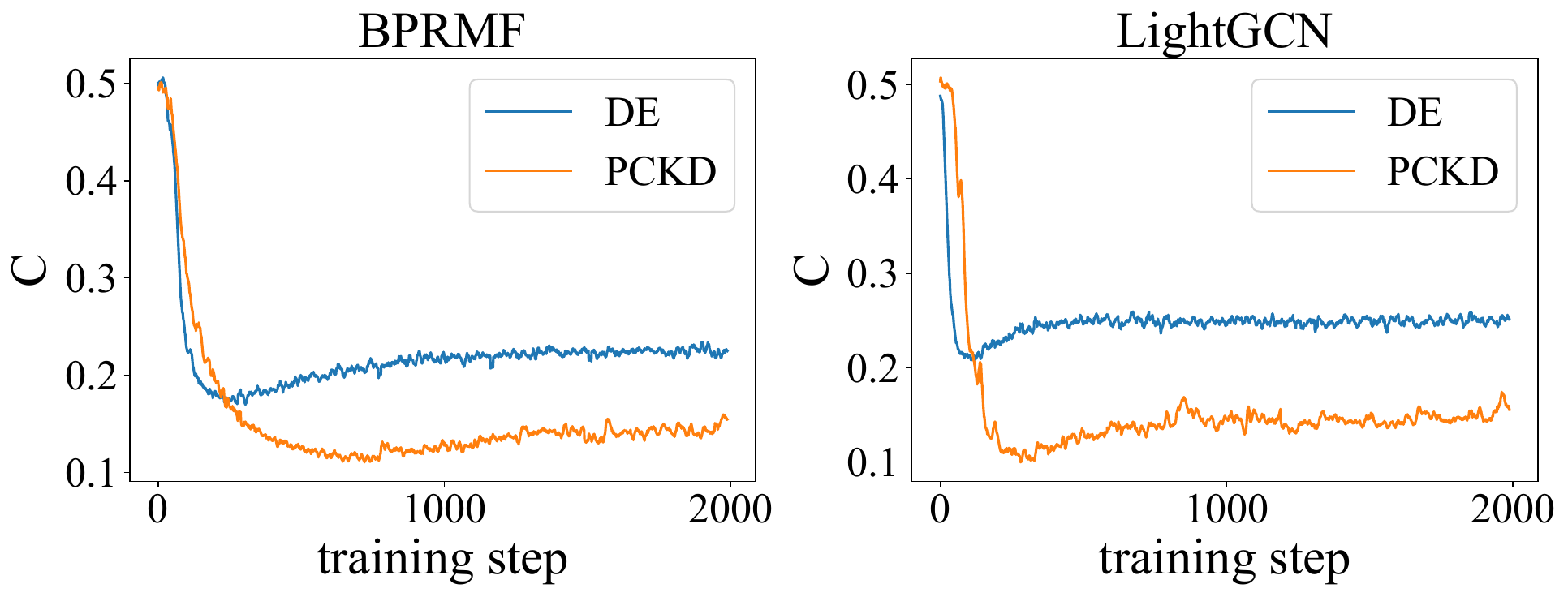}
  \caption{The dynamics of preference inconsistency as training proceeds. Experiments are conducted on CiteULike with BPRMF (left) and LightGCN (right) as the backbones.}
  \label{fig:exp1}
\end{figure}

\begin{table*}[!t]
    \caption{Ablation study. Experiments are conducted for pair-wise \ourmethod and list-wise \ourmethod separately. `w/ random sampling' stands for \ourmethod that replaces rank-aware sampling with random sampling.}
    \label{tab:abl}
    \centering
    \resizebox{\linewidth}{!}{
    \begin{tabular}{cl|cccc|cccc|cccc} \toprule
        \multirow{2}{*}{Backbone} & \multirow{2}{*}{Method} & \multicolumn{4}{c|}{CiteULike} & \multicolumn{4}{c|}{Gowalla} & \multicolumn{4}{c}{Yelp}\\
         & & R@10 & N@10 & R@20 & N@20 & R@10 & N@10 & R@20 & N@20 & R@10 & N@10 & R@20 & N@20\\
        \midrule
        \multirow{6}{*}{BPRMF} & Teacher & 0.0283 & 0.0155 & 0.0442 & 0.0198 & 0.1088 & 0.0907 & 0.1544 & 0.1053 & 0.0394 & 0.0253 & 0.0660 & 0.0339\\
         & Student & 0.0177 & 0.0098 & 0.0284 & 0.0128 & 0.0946 & 0.0820 & 0.1329 & 0.0939 & 0.0348 & 0.0222 & 0.0586 & 0.0299\\
         & DE & 0.0214 & 0.0120 & 0.0330 & 0.0147 & 0.0969 & 0.0847 & 0.1372 & 0.0971 & 0.0355 & 0.0227 & 0.0598 & 0.0305\\
         \cline{2-14}
         & \ourmethod-P & 0.0231 & 0.0133 & 0.0347 & 0.0160 & 0.0991 & 0.0859 & 0.1420 & 0.0983 & 0.0365 & 0.0235 & 0.0623 & 0.0317\\
         & w/ random sampling & 0.0222 & 0.0128 & 0.0337 & 0.0154 & 0.0975 & 0.0851 & 0.1389 & 0.0974 & 0.0358 & 0.0230 & 0.0608 & 0.0308\\
         \cline{2-14}
         & \ourmethod-L & 0.0240 & 0.0144 & 0.0355 & 0.0172 & 0.1002 & 0.0877 & 0.1445 & 0.1006 & 0.0378 & 0.0247 & 0.0633 & 0.0329\\
         & w/ random sampling & 0.0231 & 0.0136 & 0.0340 & 0.0162 & 0.0984 & 0.0858 & 0.1397 & 0.0979 & 0.0360 & 0.0232 & 0.0613 & 0.0312\\
        \midrule
        \midrule
         \multirow{6}{*}{LightGCN} & Teacher & 0.0296 & 0.0160 & 0.0461 & 0.0205 & 0.1236 & 0.1035 & 0.1730 & 0.1190 & 0.0432 & 0.0276 & 0.0716 & 0.0367\\
         & Student & 0.0215 & 0.0113 & 0.0344 & 0.0148 & 0.1098 & 0.0928 & 0.1550 & 0.1069 & 0.0363 & 0.0235 & 0.0621 & 0.0308\\
         & DE & 0.0220 & 0.0115 & 0.0349 & 0.0153 & 0.1111 & 0.0943 & 0.1578 & 0.1094 & 0.0369 & 0.0242 & 0.0641 & 0.0319\\
         \cline{2-14}
         & \ourmethod-P & 0.0227 & 0.0121 & 0.0360 & 0.0157 & 0.1141 & 0.0967 & 0.1637 & 0.1107 & 0.0376 & 0.0244 & 0.0660 & 0.0330\\
         & w/ random sampling & 0.0223 & 0.0117 & 0.0353 & 0.0155 & 0.1120 & 0.0947 & 0.1590 & 0.1097 & 0.0371 & 0.0242 & 0.0647 & 0.0322\\
         \cline{2-14}
         & \ourmethod-L & 0.0238 & 0.0128 & 0.0371 & 0.0165 & 0.1165 & 0.0991 & 0.1663 & 0.1133 & 0.0394 & 0.0249 & 0.0688 & 0.0342\\
         & w/ random sampling & 0.0226 & 0.0119 & 0.0358 & 0.0155 & 0.1123 & 0.0948 & 0.1594 & 0.1100 & 0.0372 & 0.0242 & 0.0653 & 0.0327\\
         \midrule
         \midrule
         \multirow{6}{*}{SimpleX} & Teacher & 0.0343 & 0.0191 & 0.0508 & 0.0236 & 0.1184 & 0.0943 & 0.1750 & 0.1122 & 0.0469 & 0.0303 & 0.0778 & 0.0402\\
         & Student & 0.0290 & 0.0162 & 0.0426 & 0.0199 & 0.1046 & 0.0855 & 0.1524 & 0.1006 & 0.0382 & 0.0241 & 0.0658 & 0.0330\\
         & DE & 0.0313 & 0.0165 & 0.0449 & 0.0203 & 0.1073 & 0.0869 & 0.1567 & 0.1042 & 0.0403 & 0.0253 & 0.0671 & 0.0347\\
         \cline{2-14}
         & \ourmethod-P & 0.0321 & 0.0176 & 0.0470 & 0.0220 & 0.1110 & 0.0899 & 0.1632 & 0.1064 & 0.0430 & 0.0266 & 0.0710 & 0.0358\\
         & w/ random sampling & 0.0317 & 0.0170 & 0.0458 & 0.0212 & 0.1082 & 0.0873 & 0.1579 & 0.1049 & 0.0413 & 0.0257 & 0.0690 & 0.0351\\
         \cline{2-14}
         & \ourmethod-L & 0.0331 & 0.0180 & 0.0481 & 0.0229 & 0.1139 & 0.0919 & 0.1683 & 0.1088 & 0.0439 & 0.0279 & 0.0731 & 0.0372\\
         & w/ random sampling & 0.0320 & 0.0171 & 0.0467 & 0.0219 & 0.1085 & 0.0884 & 0.01590 & 0.1058 & 0.0420 & 0.0263 & 0.0699 & 0.0359\\
         \midrule
         \midrule
         \multirow{6}{*}{JGCF} & Teacher & 0.0325 & 0.0187 & 0.0492 & 0.0243 & 0.1283 & 0.1049 & 0.1803 & 0.1230 & 0.0455 & 0.0312 & 0.0783 & 0.0412\\
         & Student & 0.0277 & 0.0129 & 0.0392 & 0.0183 & 0.1108 & 0.0943 & 0.1592 & 0.1087 & 0.0379 & 0.0251 & 0.0670 & 0.0340\\
         & DE & 0.0284 & 0.0144 & 0.0425 & 0.0195 & 0.1149 & 0.0957 & 0.1624 & 0.1114 & 0.0388 & 0.0269 & 0.0689 & 0.0360\\
         \cline{2-14}
         & \ourmethod-P & 0.0301 & 0.0150 & 0.0442 & 0.0206 & 0.1217 & 0.0997 & 0.1701 & 0.1160 & 0.0411 & 0.0285 & 0.0722 & 0.0378\\
         & w/ random sampling & 0.0291 & 0.0148 & 0.0435 & 0.0200 & 0.1201 & 0.0977 & 0.1659 & 0.1138 & 0.0396 & 0.0270 & 0.0713 & 0.0369\\
         \cline{2-14}
         & \ourmethod-L & 0.0310 & 0.0156 & 0.0451 & 0.0213 & 0.1220 & 0.1008 & 0.1723 & 0.1165 & 0.0423 & 0.0288 & 0.0743 & 0.0385\\
         & w/ random sampling & 0.0299 & 0.0150 & 0.0441 & 0.0203 & 0.1209 & 0.0982 & 0.1673 & 0.1148 & 0.0408 & 0.0271 & 0.0720 & 0.0373\\
        \bottomrule
    \end{tabular}
    }
\end{table*}

To verify that our method indeed reduces preference inconsistency, we provide the dynamics of preference inconsistency when using our method in this section. We use the list-wise regularization term since pair-wise and list-wise \ourmethod show similar trends.

In Figure~\ref{fig:exp1}, we report the dynamics of preference inconsistency with BPRMF and LightGCN as the backbones. We observe that our method significantly reduces preference inconsistency throughout training.

\begin{figure}[!t]
  \centering
  \includegraphics[width=1.0\linewidth]{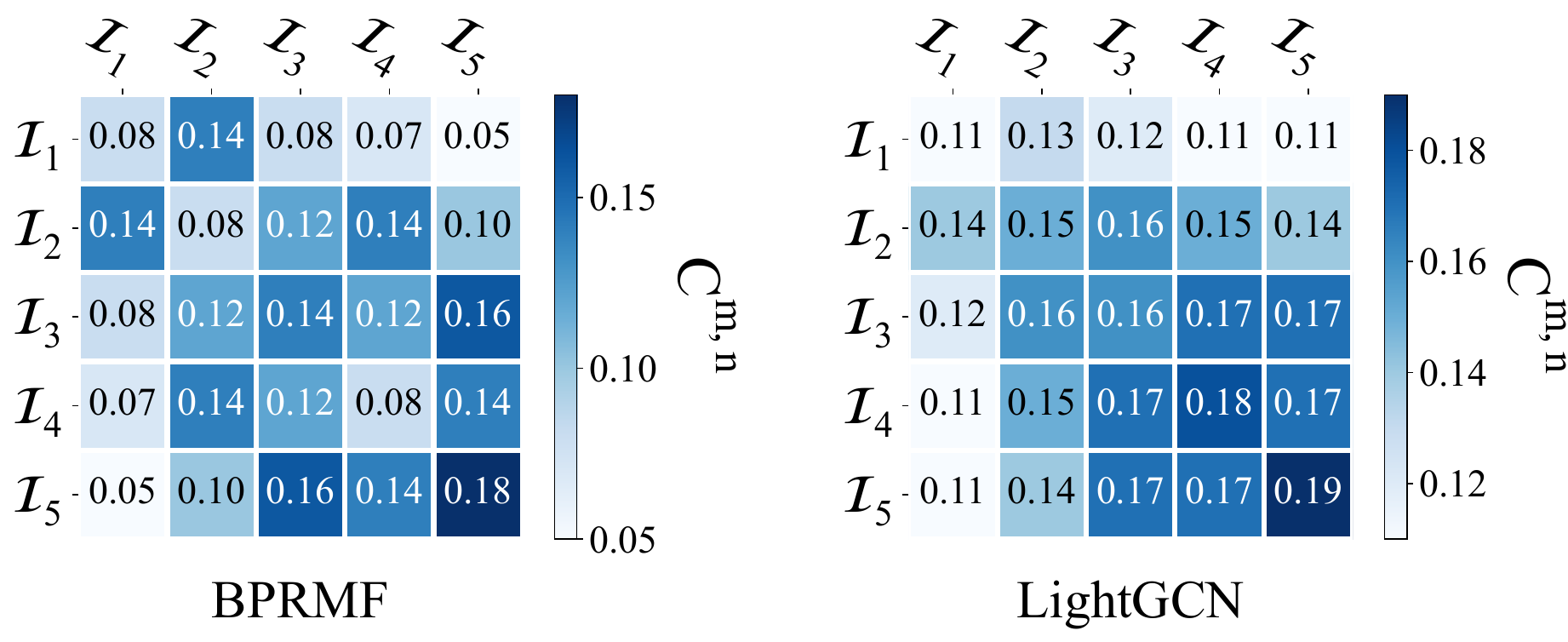}
  \caption{Group-wise preference inconsistency of \ourmethod. Experiments are conducted on CiteULike with BPRMF (left) and LightGCN (right) as the backbones.}
  \label{fig:exp2}
\end{figure}

Moreover, Figure~\ref{fig:exp2} reports the group-wise preference inconsistency. Comparing Figure~\ref{fig:exp2} with Figure~\ref{fig:method2}, it can be seen that our method significantly reduces preference inconsistency among all items, especially on items with similar preference scores. This suggests that our approach allows the model to correctly transfer user preferences even in difficult scenarios where items have similar preference scores. In addition, thanks to rank-aware sampling, preference inconsistency decreases more significantly among high-scored items. Note that this is very beneficial for knowledge distillation, as high-scored items play an important role in knowledge distillation for recommender systems.

\subsection{Ablation Study}\label{sec:abl}

In previous sections, we investigated the preference inconsistency among high-scored items and found that they are significantly affected by preference inconsistency. Combined with the fact that high-scored items are particularly crucial in knowledge distillation, we propose performing rank-aware sampling, which makes high-scored items more likely to be sampled. In this section, we conduct an ablation study to investigate the contribution of rank-aware sampling. Specifically, we replace the rank-aware sampling in pair-wise \ourmethod and list-wise \ourmethod with random sampling. The results are shown in Table~\ref{tab:abl}. We observe that:
\begin{itemize}[leftmargin=*]
    \item Even after removing rank-aware sampling, PCKD still outperforms DE. This suggests that alleviating preference inconsistency across all items equally is still effective.
    \item After replacing rank-aware sampling with random sampling, there is a significant decline in the performance, especially on Gowalla. In addition, we find that removing rank-aware sampling increases the number of items that need to be sampled, e.g., the optimal $Q$ on CiteULike increases from 10 to 50, and the optimal $Q$ increases from 30 to 200 on Gowalla. We attribute this to the fact that the proportion of items that are truly useful for knowledge distillation is small, and thus, random sampling requires increasing the sample size to find these items. Moreover, since Gowalla has many more items than the other two datasets, changes in sampling methodology have a greater impact on it.
    \item Removing rank-aware sampling results in a smaller gap between pair-wise \ourmethod and list-wise \ourmethod. We believe this is due to the greater dependence of list-wise loss on rank-aware sampling, resulting in a greater decrease in list-wise \ourmethod after removal than pair-wise \ourmethod.
\end{itemize}

\subsection{Training Efficiency}\label{sec:eff}
In this section, we report the training efficiency of our method and comparison methods. All results are obtained by testing with PyTorch on GeForce RTX 3090 GPU.

\begin{table*}[htbp]
    \caption{The comparison of the training time (seconds) per epoch. \ourmethod-P and \ourmethod-L denote the pair-wise \ourmethod and list-wise \ourmethod, respectively.}
    \label{tab:train_time}
    \centering
    \begin{tabular}{c|cccc|cccc|cccc} \toprule
        \multirow{2}{*}{Method} & \multicolumn{4}{c|}{CiteULike} & \multicolumn{4}{c|}{Gowalla} & \multicolumn{4}{c}{Yelp}\\
         & BPRMF & LightGCN & SimpleX & JGCF & BPRMF & LightGCN & SimpleX & JGCF & BPRMF & LightGCN & SimpleX & JGCF\\
        \midrule
        Student & 3.58 & 5.15 & 8.89 & 6.77 & 28.60 & 50.11 & 62.42 & 59.32 & 28.94 & 44.43 & 55.90 & 47.37\\
        \midrule
        RRD & 22.57 & 24.61 & 33.35 & 26.87 & 158.70 & 193.40 & 265.18 & 223.92 & 204.58 & 246.63 & 351.77 & 279.46\\
        UnKD & 25.15 & 34.52 & 42.48 & 37.03 & 188.32 & 215.33 & 288.72 & 247.37 & 300.15 & 288.84 & 383.12 & 311.30\\
        HTD & 10.11 & 25.62 & 38.77 & 29.93 & 65.01 & 210.37 & 277.97 & 246.38 & 78.67 & 292.79 & 278.80 & 295.92\\
        DE & 6.17 & 15.26 & 32.13 & 28.28 & 62.43 & 188.53 & 269.73 & 199.28& 52.36 & 199.80 & 244.79 & 223.04\\
        FitNet & 4.63 & 8.46 & 15.09 & 9.19 & 39.68 & 76.65 & 120.67 & 97.37 & 38.69 & 63.56 & 113.62 & 87.51\\
        \midrule
        \ourmethod-P & 8.74 & 17.97 & 34.44 & 20.38 & 91.31 & 192.08 & 273.35 & 223.40 & 75.62 & 207.69 & 261.62 & 231.41\\
        \ourmethod-L & 10.55 & 21.12 & 37.17 & 25.47 & 142.37 & 198.32 & 277.50 & 238.49 & 132.37 & 237.88 & 297.54 & 268.52\\
        \bottomrule
    \end{tabular}
\end{table*}

\begin{table*}[htbp]
    \caption{The comparison of GPU Memory (GB) required by our method and comparison methods. \ourmethod-P and \ourmethod-L denote the pair-wise \ourmethod and list-wise \ourmethod, respectively.}
    \label{tab:train_gpu}
    \centering
    \begin{tabular}{c|cccc|cccc|cccc} \toprule
        \multirow{2}{*}{Method} & \multicolumn{4}{c|}{CiteULike} & \multicolumn{4}{c|}{Gowalla} & \multicolumn{4}{c}{Yelp}\\
         & BPRMF & LightGCN & SimpleX & JGCF & BPRMF & LightGCN & SimpleX & JGCF & BPRMF & LightGCN & SimpleX & JGCF\\
        \midrule
        Student & 0.39 & 0.60 & 0.72 & 0.63 & 0.45 & 1.08 & 1.10 & 1.09 & 0.45 & 0.81 & 1.36 & 0.88\\
        \midrule
        RRD & 0.92 & 2.42 & 1.20 & 2.52 & 5.23 & 8.64 & 18.89 & 8.92 & 4.85 & 6.30 & 12.62 & 6.43\\
        UnKD & 1.22 & 2.71 & 1.52 & 2.83 & 5.50 & 8.90 & 19.77 & 10.33& 5.13 & 7.97 & 13.89 & 8.20\\
        HTD & 2.47 & 10.46 & 8.47 & 10.69 & 5.88 & 12.46 & 24.97 & 13.37 & 3.53 & 9.00 & 14.87 & 9.97\\
        DE & 0.85 & 3.54 & 3.28 & 3.60 & 1.93 & 5.68 & 8.81 & 5.93 & 0.95 & 2.69 & 5.41 & 2.71\\
        FitNet & 0.84 & 2.39 & 1.60 & 2.40 & 1.16 & 4.62 & 5.12 & 4.67 & 0.92 & 2.70 & 3.06 & 2.71\\
        \midrule
        \ourmethod-P & 0.88 & 4.02 & 3.87 & 4.07 & 2.11 & 6.42 & 10.63 & 6.50 & 2.02 & 3.79 & 7.58 & 3.97\\
        \ourmethod-L & 0.95 & 4.65 & 4.49 & 4.66 & 2.38 & 8.33 & 17.12 & 8.64 & 3.93 & 5.88 & 11.26 & 6.00\\
        \bottomrule
    \end{tabular}
\end{table*}

To investigate the training efficiency, we report the training time per epoch and the required GPU memory in Table~\ref{tab:train_time} and Table~\ref{tab:train_gpu}, respectively.
The results demonstrate that:
\begin{itemize}[leftmargin=*]
\item Our method's training cost is in between response-based and feature-based methods. Compared to existing feature-based methods, our method requires rank-aware sampling of items, increasing training costs. On the other hand, the number of items required by our method is very small, typically only 5 or 10, compared to response-based methods, which require sampling of about 200 items. We attribute this advantage to the assistance provided by the features. 
In conclusion, although our approach increases the training cost of existing feature-based methods, its final cost is still acceptable, and the resulting improvement in recommendation accuracy is significant.
\item The training cost required for pair-wise \ourmethod is much smaller than that for list-wise \ourmethod. This is consistent with our original design intent of proposing two regularization terms to enable our method to be applied to a wide range of training cost-constrained scenarios. It turns out that although pair-wise \ourmethod is not as accurate as list-wise \ourmethod in making recommendations, its low training cost allows it to be used on larger datasets. Moreover, it should be noted that the recommendation accuracy of pair-wise \ourmethod is still much higher than existing feature-based methods, so it can be used as a compromise between recommendation accuracy and training cost.
\end{itemize}

\subsection{Combination with HTD}\label{sec:htd}

\begin{figure}[htbp]
\centering
  \includegraphics[width=1.0\linewidth]{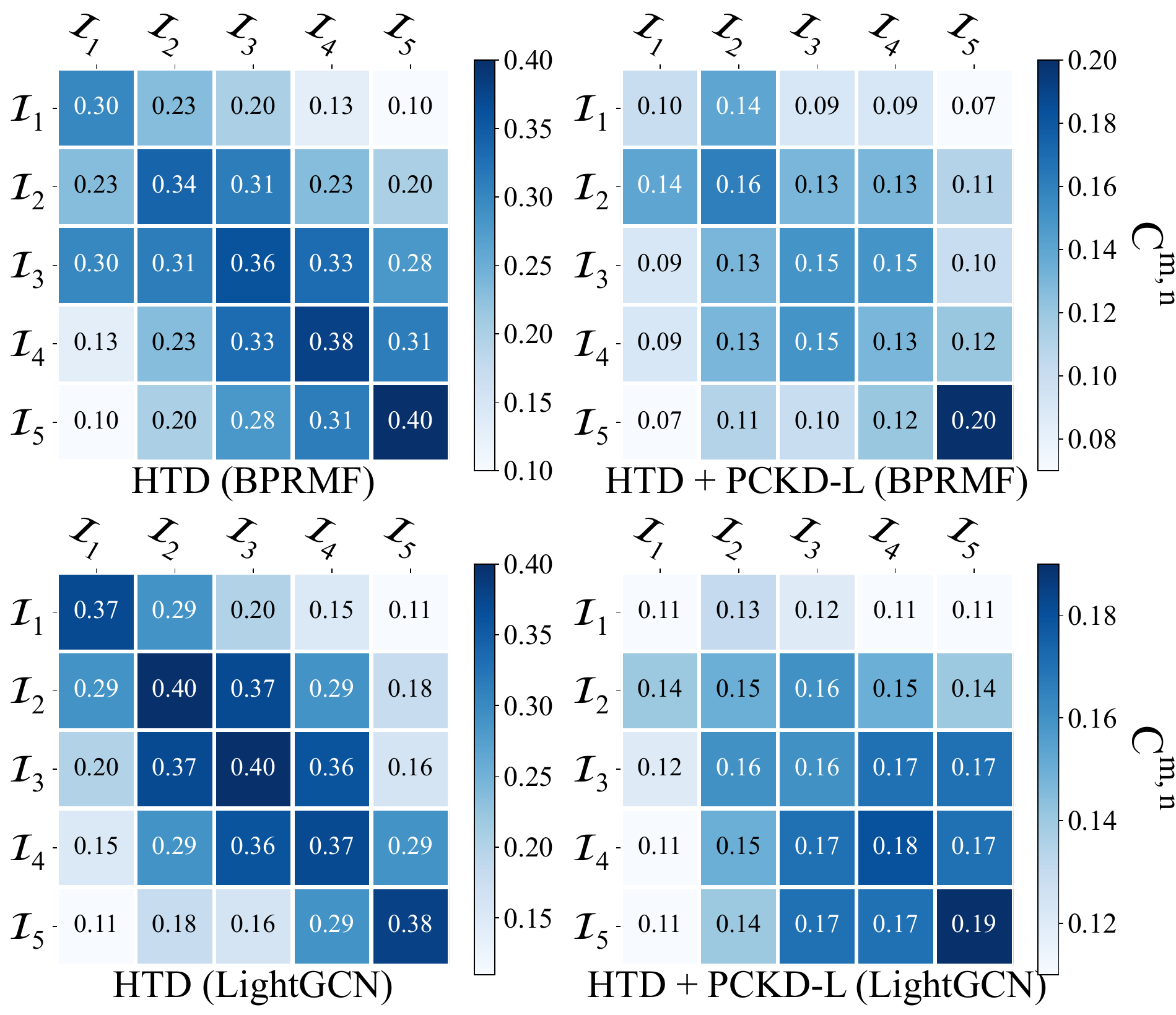}
  \caption{Group-wise preference inconsistency of original HTD (left), HTD integrating PCKD-L (right) on BPRMF (top) and LightGCN (bottom).}
  \label{fig:incon_htd_pckd}
\end{figure}

\begin{table*}[htbp]
    \caption{Performance of HTD before and after the integration of PCKD.}
    \label{tab:htd_pckd}
    \centering
    \resizebox{\linewidth}{!}{
        \begin{tabular}{cc|cccc|cccc|cccc} \toprule
        \multirow{2}{*}{Backbone} & \multirow{2}{*}{Method} & \multicolumn{4}{c|}{CiteULike} & \multicolumn{4}{c|}{Gowalla} & \multicolumn{4}{c}{Yelp}\\
         & & R@10 & N@10 & R@20 & N@20 & R@10 & N@10 & R@20 & N@20 & R@10 & N@10 & R@20 & N@20\\
        \midrule
        \multirow{4}{*}{BPRMF} & Teacher & 0.0283 & 0.0155 & 0.0442 & 0.0198 & 0.1088 & 0.0907 & 0.1544 & 0.1053 & 0.0394 & 0.0253 & 0.0660 & 0.0339\\
         & Student & 0.0177 & 0.0098 & 0.0284 & 0.0128 & 0.0946 & 0.0820 & 0.1329 & 0.0939 & 0.0348 & 0.0222 & 0.0586 & 0.0299\\
         \cline{2-14}
         & HTD & 0.0223 & 0.0127 & 0.0337 & 0.0152 & 0.0975 & 0.0852 & 0.1379 & 0.0970 & 0.0360 & 0.0231 & 0.0600 & 0.0309\\
         & HTD + PCKD-L & \textbf{0.0247} & \textbf{0.0146} & \textbf{0.0353} & \textbf{0.0178} & \textbf{0.1001} & \textbf{0.881} & \textbf{0.1448} & \textbf{0.1001} & \textbf{0.0373} & \textbf{0.0246} & \textbf{0.0633} & \textbf{0.0331}\\
        \midrule
        \midrule
         \multirow{4}{*}{LightGCN} & Teacher & 0.0296 & 0.0160 & 0.0461 & 0.0205 & 0.1236 & 0.1035 & 0.1730 & 0.1190 & 0.0432 & 0.0276 & 0.0716 & 0.0367\\
         & Student & 0.0215 & 0.0113 & 0.0344 & 0.0148 & 0.1098 & 0.0928 & 0.1550 & 0.1069 & 0.0363 & 0.0235 & 0.0621 & 0.0308\\
         \cline{2-14}
         & HTD & 0.0225 & 0.0120 & 0.0352 & 0.0154 & 0.1139 & 0.0951 & 0.1589 & 0.1102 & 0.0375 & 0.0245 & 0.0653 & 0.0330\\
         & HTD + PCKD-L & \textbf{0.0241} & \textbf{0.0134} & \textbf{0.0369} & \textbf{0.0166} & \textbf{0.1172} & \textbf{0.0999} & \textbf{0.1660} & \textbf{0.1137} & \textbf{0.0400} & \textbf{0.0247} & \textbf{0.0693} & \textbf{0.0344}\\
         \midrule
         \midrule
         \multirow{4}{*}{SimpleX} & Teacher & 0.0343 & 0.0191 & 0.0508 & 0.0236 & 0.1184 & 0.0943 & 0.1750 & 0.1122 & 0.0469 & 0.0303 & 0.0778 & 0.0402\\
         & Student & 0.0290 & 0.0162 & 0.0426 & 0.0199 & 0.1046 & 0.0855 & 0.1524 & 0.1006 & 0.0382 & 0.0241 & 0.0658 & 0.0330\\
         \cline{2-14}
         & HTD & 0.0317 & 0.0172 & 0.0462 & 0.0215 & 0.1089 & 0.0883 & 0.1599 & 0.1057 & 0.0406 & 0.0258 & 0.0690 & 0.0353\\
         & HTD + PCKD-L & \textbf{0.0330} & \textbf{0.0180} & \textbf{0.0484} & \textbf{0.0230} & \textbf{0.1142} & \textbf{0.0928} & \textbf{0.1692} & \textbf{0.1083} & \textbf{0.0438} & \textbf{0.0281} & \textbf{0.0728} & \textbf{0.0374}\\
         \midrule
         \midrule
         \multirow{4}{*}{JGCF} & Teacher & 0.0325 & 0.0187 & 0.0492 & 0.0243 & 0.1283 & 0.1049 & 0.1803 & 0.1230 & 0.0455 & 0.0312 & 0.0783 & 0.0412\\
         & Student & 0.0277 & 0.0129 & 0.0392 & 0.0183 & 0.1108 & 0.0943 & 0.1592 & 0.1087 & 0.0379 & 0.0251 & 0.0670 & 0.0340\\
         \cline{2-14}
         & HTD & 0.0289 & 0.0145 & 0.0430 & 0.0198 & 0.1148 & 0.0969 & 0.1640 & 0.1118 & 0.0390 & 0.0270 & 0.0687 & 0.0368\\
         & HTD + PCKD-L & \textbf{0.0311} & \textbf{0.0153} & \textbf{0.0442} & \textbf{0.0210} & \textbf{0.1223} & \textbf{0.1010} & \textbf{0.1719} & \textbf{0.1160} & \textbf{0.0427} & \textbf{0.0288} & \textbf{0.0737} & \textbf{0.0391}\\
        \bottomrule
        \end{tabular}
    }
\end{table*}

HTD~\cite{kang2021topology} is a relation-based knowledge distillation method. It proposes to distill the sample relation hierarchically to alleviate the large capacity gap between the student and the teacher. Although HTD is not a feature-based approach, it is similar to DE in that it uses an expert module comprising multiple projectors and a selection network to cluster users and items. This allows our proposed regularization terms for projectors to be used in HTD.

Firstly, we plot the existence of preference inconsistency in HTD in the left subplot of Figure~\ref{fig:incon_htd_pckd}. It demonstrates a serious preference inconsistency problem in HTD, which is even worse than in DE. Then, in the right two subplots of Figure~\ref{fig:incon_htd_pckd}, we show the preference inconsistency after combining our proposed PCKD-L with HTD. As we can see, adding our proposed regularity term significantly reduces the preference inconsistency in HTD. Moreover, to verify that decreasing the preference inconsistency in HTD can lead to recommendation performance improvement, we compare the performance of HTD before and after adding PCKD-L in Table~\ref{tab:htd_pckd}. The results demonstrate that the performance of HTD is significantly improved after adding PCKD-L. This shows that our method can be widely used in scenarios where projectors are used, even in non-feature-based methods.

\subsection{Comparison with Response-based Method}\label{sec:response}

\begin{figure}[htbp]
\centering
  \includegraphics[width=1.0\linewidth]{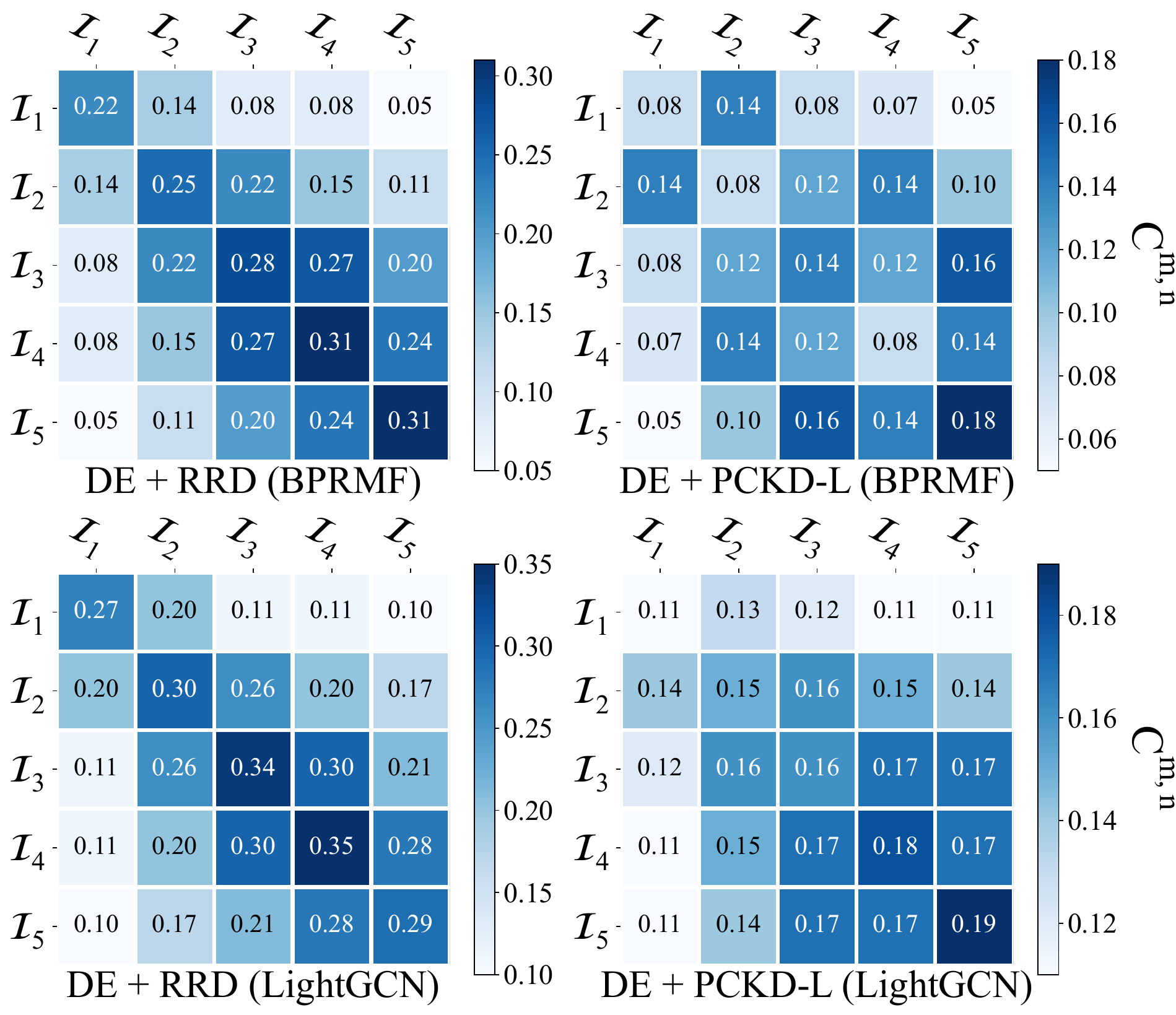}
  \caption{Group-wise preference inconsistency of combining DE with RRD (left) and with our proposed PCKD-L (right) on BPRMF (top) and LightGCN (bottom).}
  \label{fig:incon_rrd_pckd}
\end{figure}

Response-based knowledge distillation methods improve the student model by enabling it to learn the predictive distribution of the teacher. Recent works demonstrate that combining feature-based methods with response-based methods can lead to greater improvement~\cite{kang2020rrd,cui2024distillation}. In addition, since the response-based approach brings the predicted distributions of the teacher and student closer, there will be the possibility of reducing preference inconsistency. This section empirically shows that it is far less effective at lowering preference inconsistency than our method.

We based our experiments on RRD~\cite{kang2020rrd}, a widely used response-based method. In Fig.~\ref{fig:incon_rrd_pckd}, we show the preference inconsistency for combining DE with RRD (left) and with our proposed PCKD-L (right). The results demonstrate that combining response-based distillation does reduce preference inconsistency, but to a small extent, much less than our proposed PCKD. We believe that this is because our approach directly constrains the projector, whereas the response-based approach can only indirectly reduce the emergence of erroneous projectors by influencing the student's predictive user preference.

\subsection{Performance of \ourmethod-H}\label{sec:pckd-h}

\begin{table*}[htbp]
    \caption{Performance of different methods combining pair-wise and list-wise \ourmethod.}
    \label{tab:pckd-h}
    \centering
    \resizebox{\linewidth}{!}{
        \begin{tabular}{cc|cccc|cccc|cccc} \toprule
        \multirow{2}{*}{Backbone} & \multirow{2}{*}{Method} & \multicolumn{4}{c|}{CiteULike} & \multicolumn{4}{c|}{Gowalla} & \multicolumn{4}{c}{Yelp}\\
         & & R@10 & N@10 & R@20 & N@20 & R@10 & N@10 & R@20 & N@20 & R@10 & N@10 & R@20 & N@20\\
        \midrule
        \multirow{4}{*}{BPRMF} & \ourmethod-P & 0.0231 & 0.0133 & 0.0347 & 0.0160 & 0.0991 & 0.0859 & 0.1420 & 0.0983 & 0.0365 & 0.0235 & 0.0623 & 0.0317\\
         & \ourmethod-L & 0.0240 & 0.0144 & 0.0355 & 0.0172 & 0.1002 & 0.0877 & 0.1445 & 0.1006 & 0.0378 & 0.0247 & 0.0633 & 0.0329\\
         \cline{2-14}
         & Vanilla & 0.0241 & 0.0140 & 0.0357 & 0.0177 & 0.1000 & 0.0871 & 0.1447 & 0.1006 & 0.0377 & 0.0239 & 0.0628 & 0.0331\\
         & \ourmethod-H & \textbf{0.0250} & \textbf{0.0149} & \textbf{0.0368} & \textbf{0.0182} & \textbf{0.1014} & \textbf{0.0892} & \textbf{0.1473} & \textbf{0.1020} & \textbf{0.0383} & \textbf{0.0249} & \textbf{0.0651} & \textbf{0.0334}\\
        \midrule
        \midrule
         \multirow{4}{*}{LightGCN} & \ourmethod-P & 0.0227 & 0.0121 & 0.0360 & 0.0157 & 0.1141 & 0.0967 & 0.1637 & 0.1107 & 0.0376 & 0.0244 & 0.0660 & 0.0330\\
         & \ourmethod-L & 0.0238 & 0.0128 & 0.0371 & 0.0165 & 0.1165 & 0.0991 & 0.1663 & 0.1133 & 0.0394 & 0.0249 & 0.0688 & 0.0342\\
         \cline{2-14}
         & Vanilla & 0.0236 & 0.0126 & 0.0373 & 0.0169 & 0.1166 & 0.0983 & 0.1559 & 0.1120 & 0.0390 & 0.0249 & 0.0692 & 0.0339\\
         & \ourmethod-H & \textbf{0.0249} & \textbf{0.0140} & \textbf{0.0381} & \textbf{0.0181} & \textbf{0.1173} & \textbf{0.1002} & \textbf{0.1681} & \textbf{0.1157} & \textbf{0.0407} & \textbf{0.0255} & \textbf{0.0680} & \textbf{0.0351}\\
         \midrule
         \midrule
         \multirow{4}{*}{SimpleX} & \ourmethod-P & 0.0321 & 0.0176 & 0.0470 & 0.0220 & 0.1110 & 0.0899 & 0.1632 & 0.1064 & 0.0430 & 0.0266 & 0.0710 & 0.0358\\
         & \ourmethod-L & 0.0331 & 0.0180 & 0.0481 & 0.0229 & 0.1139 & 0.0919 & 0.1683 & 0.1088 & 0.0439 & 0.0279 & 0.0731 & 0.0372\\
         \cline{2-14}
         & Vanilla & 0.0333 & 0.0181 & 0.0479 & 0.0221 & 0.1129 & 0.0923 & 0.1688 & 0.1088 & 0.0436 & 0.0280 & 0.0735 & 0.0373\\
         & \ourmethod-H & \textbf{0.0339} & \textbf{0.0187} & \textbf{0.0490} & \textbf{0.0233} & \textbf{0.1152} & \textbf{0.0930} & \textbf{0.1701} & \textbf{0.1099} & \textbf{0.0445} & \textbf{0.0283} & \textbf{0.0743} & \textbf{0.0382}\\
         \midrule
         \midrule
         \multirow{4}{*}{JGCF} & PCKD-P & 0.0301 & 0.0150 & 0.0442 & 0.0206 & 0.1217 & 0.0997 & 0.1701 & 0.1160 & 0.0411 & 0.0285 & 0.0722 & 0.0378\\
         & PCKD-L & 0.0310 & 0.0156 & 0.0451 & 0.0213 & 0.1220 & 0.1008 & 0.1723 & 0.1165 & 0.0423 & 0.0288 & 0.0743 & 0.0385\\
         \cline{2-14}
         & Vanilla & 0.0303 & 0.0157 & 0.0455 & 0.0215 & 0.1222 & 0.1011 & 0.1725 & 0.1166 & 0.0420 & 0.0287 & 0.0739 & 0.0382\\
         & \ourmethod-H & \textbf{0.0317} & \textbf{0.0167} & \textbf{0.0462} & \textbf{0.0230} & \textbf{0.1237} & \textbf{0.1015} & \textbf{0.1740} & \textbf{0.1171} & \textbf{0.0430} & \textbf{0.0292} & \textbf{0.0755} & \textbf{0.0399}\\
        \bottomrule
        \end{tabular}
    }
\end{table*}

While we designed PCKD-P and PCKD-L to be used in scenarios with different training budgets, we also proposed PCKD-H to further improve recommendation performance by combining PCKD-P and PCKD-L. Before combining the two regularization terms, we propose to modify the sampling strategy in PCKD-P to make the two regularization terms complement each other. In this section, we empirically validate the effectiveness of our approach.

In Table~\ref{tab:pckd-h}, we report the performance of our proposed PCKD-H. We also report the performance when we do not modify the sampling strategy of PCKD-P, but combine the two regularization terms directly through the hyperparameter $\alpha$, which is denoted as Vanilla.
We find that PCKD-H outperforms PCKD-P and PCKD-L, suggesting that combining the two losses can further improve the recommended performance. Moreover, PCKD-H significantly outperforms Vanilla. The results demonstrate the effectiveness of our proposed method of combining the two losses.

\subsection{Hyperparameter Analysis}\label{sec:hyper}

\begin{figure}[htbp]
  \centering
  \includegraphics[width=1.0\linewidth]{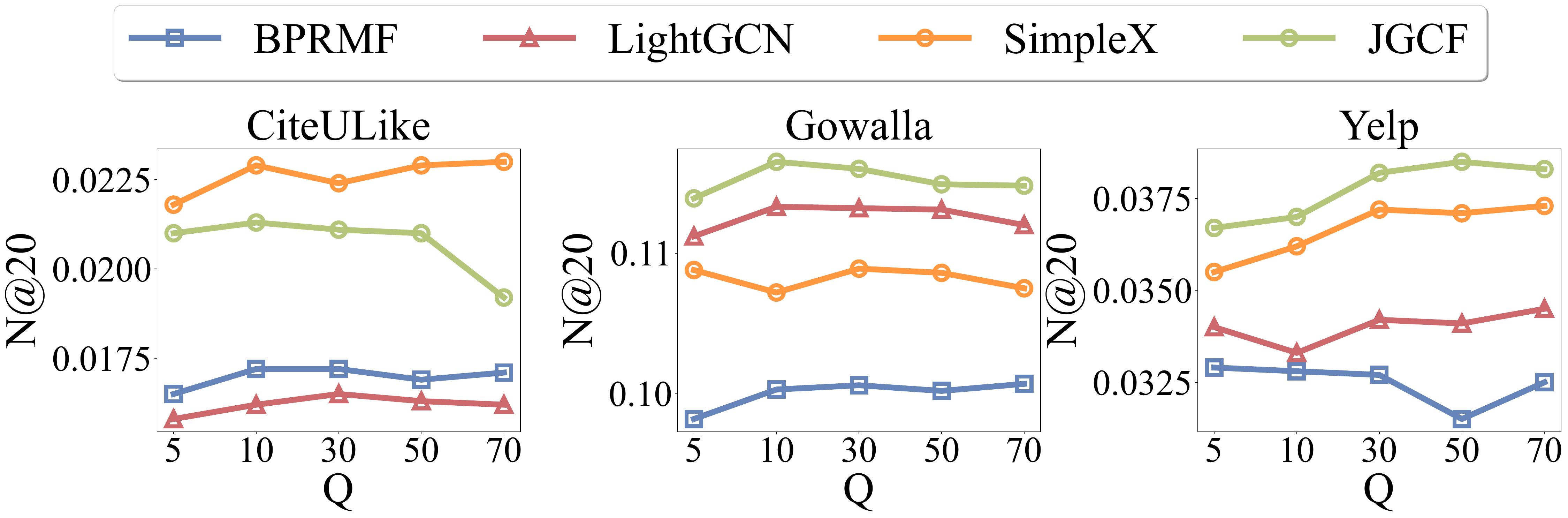}
  \caption{Effects of $Q$.}
  \label{fig:hyper_Q}
\end{figure}

\begin{figure}[htbp]
  \centering
  \includegraphics[width=1.0\linewidth]{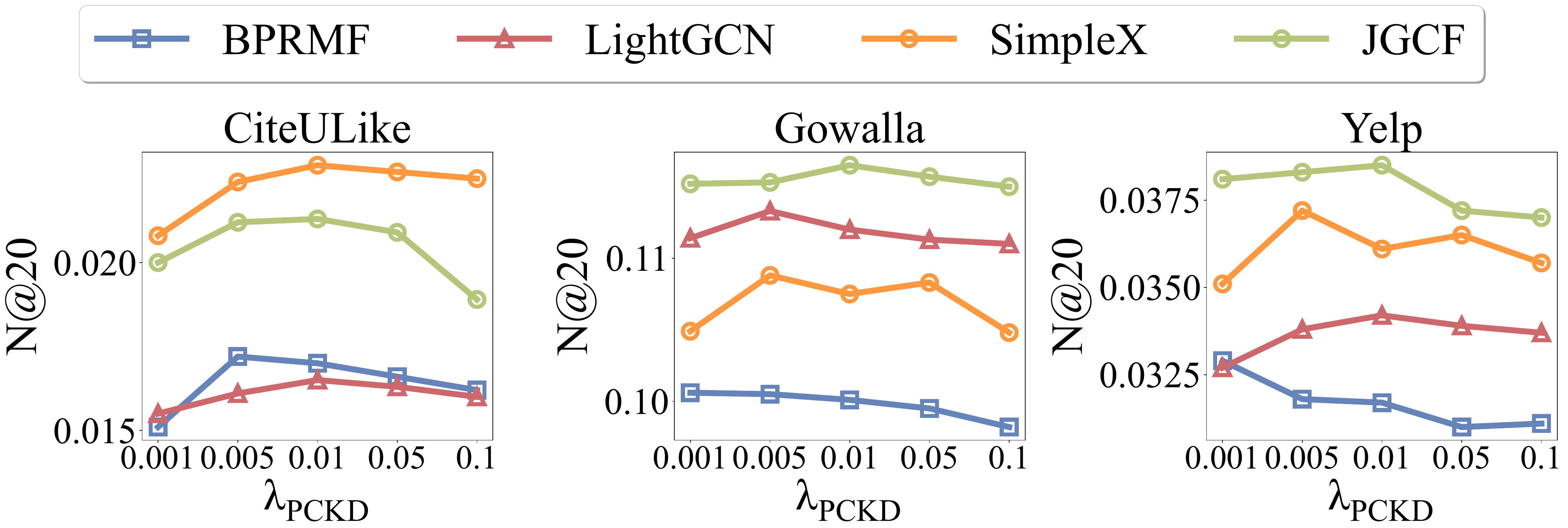}
  \caption{Effects of $\lambda_{\ourmethod}$.}
  \label{fig:hyper_lambda}
\end{figure}

\begin{figure}[htbp]
  \centering
  \includegraphics[width=1.0\linewidth]{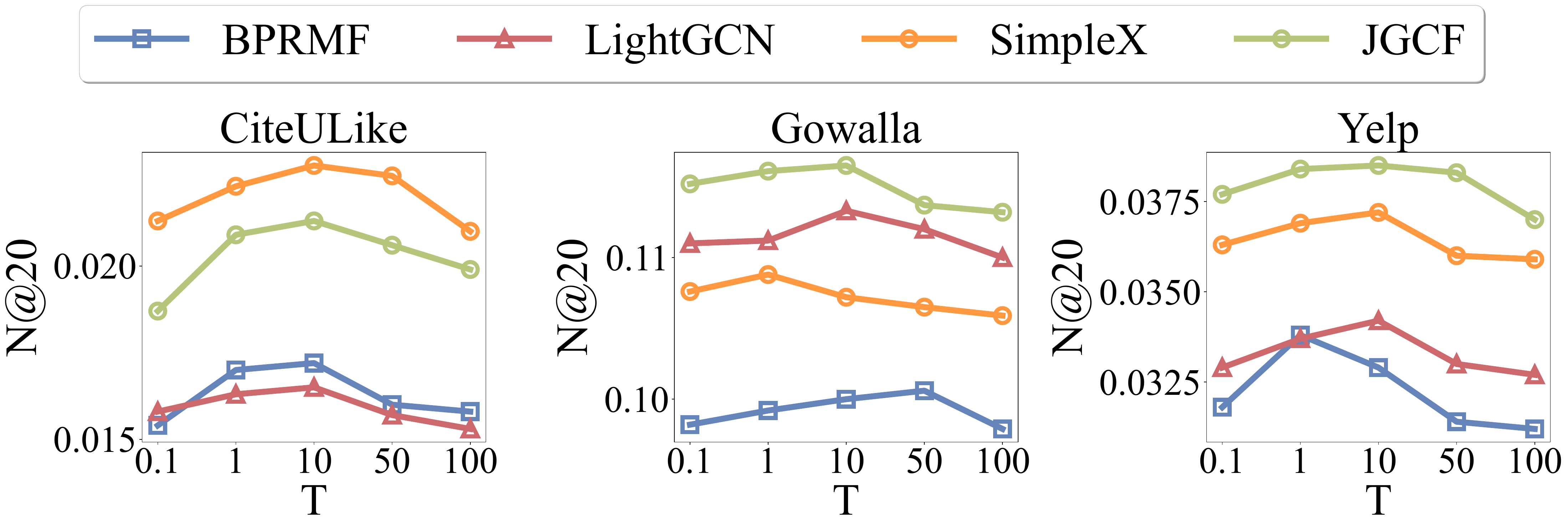}
  \caption{Effects of $T$.}
  \label{fig:hyper_T}
\end{figure}

This section analyses the effects of hyperparameters of \ourmethod regarding NDCG@20.

\textbf{Effects of $Q$.}
In list-wise \ourmethod, we sample $Q$ items independently for each user to construct $\mathcal{Q}_{u}$. The effects of $Q$ are reported in Figure~\ref{fig:hyper_Q}. The results show that setting $Q$ to some very small values, such as 10, can already give good results. On the other hand, taking a larger value of $Q$, although it can bring further improvement in some cases, is not worth it compared to the increased training cost it brings. Therefore, we recommend setting the value of $Q$ to $10$.

\textbf{Effects of $\lambda_{\ourmethod}$.}
In Eq.(\ref{eq:all}), hyperparameter $\lambda_{\ourmethod}$ represents the weight of $\mathcal{L}_{\ourmethod}$. The effects of $\lambda_{\ourmethod}$ are shown in Figure~\ref{fig:hyper_lambda}. Because the types of loss functions of the proposed methods differ from those of the base models, it is important to balance the losses by properly setting $\beta$. Our experiments find that setting $\lambda_{\ourmethod}$ to $0.005$ usually resulted in good performance.

\textbf{Effects of $T$.}
In pair-wise \ourmethod and list-wise \ourmethod, we perform rank-aware sampling to emphasize the high-scored items. Hyperparameter $T$ in Eq.(\ref{eq:sample}) controls the probability of high-scored items being sampled. Figure~\ref{fig:hyper_T} shows the effects of $T$. On the one hand, a $T$ that is too small can cause the PCKD to focus only on high-scored items and ignore preference inconsistencies in too many of the remaining items. On the other hand, a $T$ that is too large can result in the importance of high-scored items not being highlighted. Therefore, the optimal $T$ values range from 1 to 10.

\begin{figure}[!t]
  \centering
  \includegraphics[width=1.0\linewidth]{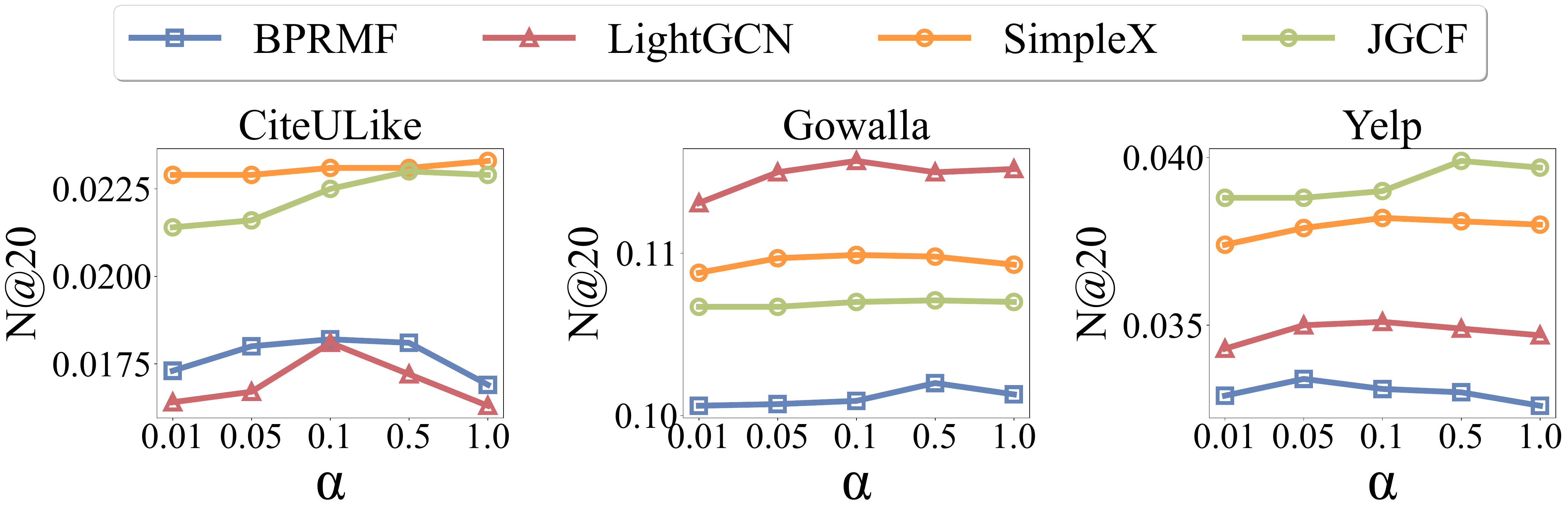}
  \caption{Effects of $\alpha$.}
  \label{fig:hyper_alpha}
\end{figure}

\textbf{Effects of $\alpha$.}
In the hybrid method, i.e., \ourmethod-H in \ref{eq:loss_comb}, we use $\alpha$ to combine the pair-wise and list-wise losses. Figure~\ref{fig:hyper_alpha} shows the effects of $\alpha$. Since the two losses are responsible for different items, it is important to balance them. We empirically find that the optimal $\alpha$ values range from 0.1 to 1.0.

\section{Conclusion}

In this paper, we focus on feature-based knowledge distillation for recommender systems and investigate the preference inconsistency caused by projectors. We show that user preferences derived from projected features are highly likely to differ from those derived from student features. This wastes the power of feature-based distillation because user preferences are what we really need in recommender systems. Therefore, we propose two regularization terms to alleviate preference inconsistency: pair-wise \ourmethod and list-wise \ourmethod. They focus on items with high preference scores and significantly mitigate preference inconsistency. They can be used in scenarios with different training cost constraints. Moreover, we propose a hybrid method to combine these two regularization terms. The experimental results over three public datasets and three backbones demonstrate the effectiveness of \ourmethod.

Although list-wise \ourmethod can greatly reduce preference inconsistency, it requires a larger training cost than pair-wise PCKD and existing feature-based knowledge distillation methods. Therefore, readers have to make a trade-off between recommendation accuracy and training cost, which is a limitation of our approach. Consequently, we are committed to reducing preference inconsistency without increasing the training cost in future work.

\bibliographystyle{abbrv}
\bibliography{IEEEabrv,refs.bib}

\begin{thebibliography}{10}

\bibitem{chen2023unbiased}
G.~Chen, J.~Chen, F.~Feng, S.~Zhou, and X.~He.
\newblock Unbiased knowledge distillation for recommendation.
\newblock In {\em Proceedings of the Sixteenth ACM International Conference on Web Search and Data Mining}, pages 976--984, 2023.

\bibitem{chen2020learning}
H.~Chen, Y.~Wang, C.~Xu, C.~Xu, and D.~Tao.
\newblock Learning student networks via feature embedding.
\newblock {\em IEEE Transactions on Neural Networks and Learning Systems}, 32(1):25--35, 2020.

\bibitem{chen2022improved}
Y.~Chen, S.~Wang, J.~Liu, X.~Xu, F.~de~Hoog, and Z.~Huang.
\newblock Improved feature distillation via projector ensemble.
\newblock {\em Advances in Neural Information Processing Systems}, 35:12084--12095, 2022.

\bibitem{cho2011friendship}
E.~Cho, S.~A. Myers, and J.~Leskovec.
\newblock Friendship and mobility: user movement in location-based social networks.
\newblock In {\em Proceedings of the 17th ACM SIGKDD international conference on Knowledge discovery and data mining}, pages 1082--1090, 2011.

\bibitem{cui2024distillation}
Y.~Cui, F.~Liu, P.~Wang, B.~Wang, H.~Tang, Y.~Wan, J.~Wang, and J.~Chen.
\newblock Distillation matters: Empowering sequential recommenders to match the performance of large language models.
\newblock In {\em Proceedings of the 18th ACM Conference on Recommender Systems}, pages 507--517, 2024.

\bibitem{gou2021knowledge}
J.~Gou, B.~Yu, S.~J. Maybank, and D.~Tao.
\newblock Knowledge distillation: A survey.
\newblock {\em International Journal of Computer Vision}, 129:1789--1819, 2021.

\bibitem{guo2023manipulating}
J.~Guo, L.~Du, X.~Chen, X.~Ma, Q.~Fu, S.~Han, D.~Zhang, and Y.~Zhang.
\newblock On manipulating signals of user-item graph: A jacobi polynomial-based graph collaborative filtering.
\newblock In {\em Proceedings of the 29th ACM SIGKDD Conference on Knowledge Discovery and Data Mining}, pages 602--613, 2023.

\bibitem{he2020lightgcn}
X.~He, K.~Deng, X.~Wang, Y.~Li, Y.~Zhang, and M.~Wang.
\newblock Lightgcn: Simplifying and powering graph convolution network for recommendation.
\newblock In {\em Proceedings of the 43rd International ACM SIGIR conference on research and development in Information Retrieval}, pages 639--648, 2020.

\bibitem{hinton2015distilling}
G.~Hinton, O.~Vinyals, and J.~Dean.
\newblock Distilling the knowledge in a neural network.
\newblock {\em arXiv preprint arXiv:1503.02531}, 2015.

\bibitem{huang2023aligning}
F.~Huang, Z.~Wang, X.~Huang, Y.~Qian, Z.~Li, and H.~Chen.
\newblock Aligning distillation for cold-start item recommendation.
\newblock In {\em Proceedings of the 46th International ACM SIGIR Conference on Research and Development in Information Retrieval}, pages 1147--1157, 2023.

\bibitem{huang2022knowledge}
T.~Huang, S.~You, F.~Wang, C.~Qian, and C.~Xu.
\newblock Knowledge distillation from a stronger teacher.
\newblock {\em Advances in Neural Information Processing Systems}, 35:33716--33727, 2022.

\bibitem{huang2022masked}
T.~Huang, Y.~Zhang, S.~You, F.~Wang, C.~Qian, J.~Cao, and C.~Xu.
\newblock Masked distillation with receptive tokens.
\newblock {\em arXiv preprint arXiv:2205.14589}, 2022.

\bibitem{kang2020rrd}
S.~Kang, J.~Hwang, W.~Kweon, and H.~Yu.
\newblock De-rrd: A knowledge distillation framework for recommender system.
\newblock In {\em Proceedings of the 29th ACM International Conference on Information \& Knowledge Management}, pages 605--614, 2020.

\bibitem{kang2021item}
S.~Kang, J.~Hwang, W.~Kweon, and H.~Yu.
\newblock Item-side ranking regularized distillation for recommender system.
\newblock {\em Information Sciences}, 580:15--34, 2021.

\bibitem{kang2021topology}
S.~Kang, J.~Hwang, W.~Kweon, and H.~Yu.
\newblock Topology distillation for recommender system.
\newblock In {\em Proceedings of the 27th ACM SIGKDD Conference on Knowledge Discovery \& Data Mining}, pages 829--839, 2021.

\bibitem{kang2023distillation}
S.~Kang, W.~Kweon, D.~Lee, J.~Lian, X.~Xie, and H.~Yu.
\newblock Distillation from heterogeneous models for top-k recommendation.
\newblock In {\em Proceedings of the ACM Web Conference 2023}, pages 801--811, 2023.

\bibitem{kang2022personalized}
S.~Kang, D.~Lee, W.~Kweon, and H.~Yu.
\newblock Personalized knowledge distillation for recommender system.
\newblock {\em Knowledge-Based Systems}, 239:107958, 2022.

\bibitem{klenitskiy2023turning}
A.~Klenitskiy and A.~Vasilev.
\newblock Turning dross into gold loss: is bert4rec really better than sasrec?
\newblock In {\em Proceedings of the 17th ACM Conference on Recommender Systems}, pages 1120--1125, 2023.

\bibitem{koren2009matrix}
Y.~Koren, R.~Bell, and C.~Volinsky.
\newblock Matrix factorization techniques for recommender systems.
\newblock {\em Computer}, 42(8):30--37, 2009.

\bibitem{kweon2021bidirectional}
W.~Kweon, S.~Kang, and H.~Yu.
\newblock Bidirectional distillation for top-k recommender system.
\newblock In {\em Proceedings of the Web Conference 2021}, pages 3861--3871, 2021.

\bibitem{lee2019collaborative}
J.-w. Lee, M.~Choi, J.~Lee, and H.~Shim.
\newblock Collaborative distillation for top-n recommendation.
\newblock In {\em 2019 IEEE International Conference on Data Mining (ICDM)}, pages 369--378. IEEE, 2019.

\bibitem{lee2021dual}
Y.~Lee and K.-E. Kim.
\newblock Dual correction strategy for ranking distillation in top-n recommender system.
\newblock In {\em Proceedings of the 30th ACM International Conference on Information \& Knowledge Management}, pages 3186--3190, 2021.

\bibitem{liang2018variational}
D.~Liang, R.~G. Krishnan, M.~D. Hoffman, and T.~Jebara.
\newblock Variational autoencoders for collaborative filtering.
\newblock In {\em Proceedings of the 2018 world wide web conference}, pages 689--698, 2018.

\bibitem{liu2023norm}
X.~Liu, L.~Li, C.~Li, and A.~Yao.
\newblock Norm: Knowledge distillation via n-to-one representation matching.
\newblock {\em arXiv preprint arXiv:2305.13803}, 2023.

\bibitem{mao2021simplex}
K.~Mao, J.~Zhu, J.~Wang, Q.~Dai, Z.~Dong, X.~Xiao, and X.~He.
\newblock Simplex: A simple and strong baseline for collaborative filtering.
\newblock In {\em Proceedings of the 30th ACM International Conference on Information \& Knowledge Management}, pages 1243--1252, 2021.

\bibitem{ohsaka2023curse}
N.~Ohsaka and R.~Togashi.
\newblock Curse of" low" dimensionality in recommender systems.
\newblock {\em arXiv preprint arXiv:2305.13597}, 2023.

\bibitem{park2019relational}
W.~Park, D.~Kim, Y.~Lu, and M.~Cho.
\newblock Relational knowledge distillation.
\newblock In {\em Proceedings of the IEEE/CVF conference on computer vision and pattern recognition}, pages 3967--3976, 2019.

\bibitem{passalis2018learning}
N.~Passalis and A.~Tefas.
\newblock Learning deep representations with probabilistic knowledge transfer.
\newblock In {\em Proceedings of the European Conference on Computer Vision (ECCV)}, pages 268--284, 2018.

\bibitem{passalis2020probabilistic}
N.~Passalis, M.~Tzelepi, and A.~Tefas.
\newblock Probabilistic knowledge transfer for lightweight deep representation learning.
\newblock {\em IEEE Transactions on Neural Networks and Learning Systems}, 32(5):2030--2039, 2020.

\bibitem{ren2021unbiased}
Y.~Ren, H.~Tang, and S.~Zhu.
\newblock Unbiased pairwise learning to rank in recommender systems.
\newblock {\em arXiv preprint arXiv:2111.12929}, 2021.

\bibitem{rendle2014improving}
S.~Rendle and C.~Freudenthaler.
\newblock Improving pairwise learning for item recommendation from implicit feedback.
\newblock In {\em Proceedings of the 7th ACM international conference on Web search and data mining}, pages 273--282, 2014.

\bibitem{rendle2012bpr}
S.~Rendle, C.~Freudenthaler, Z.~Gantner, and L.~Schmidt-Thieme.
\newblock Bpr: Bayesian personalized ranking from implicit feedback.
\newblock {\em arXiv preprint arXiv:1205.2618}, 2012.

\bibitem{romero2014fitnets}
A.~Romero, N.~Ballas, S.~E. Kahou, A.~Chassang, C.~Gatta, and Y.~Bengio.
\newblock Fitnets: Hints for thin deep nets.
\newblock {\em arXiv preprint arXiv:1412.6550}, 2014.

\bibitem{shi2023quantize}
L.~Shi, Y.~Liu, J.~Wang, and W.~Zhang.
\newblock Quantize sequential recommenders without private data.
\newblock In {\em Proceedings of the ACM Web Conference 2023}, pages 1043--1052, 2023.

\bibitem{tang2018ranking}
J.~Tang and K.~Wang.
\newblock Ranking distillation: Learning compact ranking models with high performance for recommender system.
\newblock In {\em Proceedings of the 24th ACM SIGKDD international conference on knowledge discovery \& data mining}, pages 2289--2298, 2018.

\bibitem{wang2013collaborative}
H.~Wang, B.~Chen, and W.-J. Li.
\newblock Collaborative topic regression with social regularization for tag recommendation.
\newblock In {\em IJCAI}, volume~13, pages 2719--2725, 2013.

\bibitem{wang2019neural}
X.~Wang, X.~He, M.~Wang, F.~Feng, and T.-S. Chua.
\newblock Neural graph collaborative filtering.
\newblock In {\em Proceedings of the 42nd international ACM SIGIR conference on Research and development in Information Retrieval}, pages 165--174, 2019.

\bibitem{xu2023stablegcn}
C.~Xu, J.~Wang, and W.~Zhang.
\newblock Stablegcn: Decoupling and reconciling information propagation for collaborative filtering.
\newblock {\em IEEE Transactions on Knowledge and Data Engineering}, 2023.

\bibitem{xu2024fairly}
C.~Xu, Z.~Zhu, J.~Wang, J.~Wang, and W.~Zhang.
\newblock Fairly evaluating large language model-based recommendation needs revisit the cross-entropy loss.
\newblock {\em arXiv preprint arXiv:2402.06216}, 2024.

\bibitem{yang2022masked}
Z.~Yang, Z.~Li, M.~Shao, D.~Shi, Z.~Yuan, and C.~Yuan.
\newblock Masked generative distillation.
\newblock In {\em European Conference on Computer Vision}, pages 53--69. Springer, 2022.

\bibitem{zagoruyko2016paying}
S.~Zagoruyko and N.~Komodakis.
\newblock Paying more attention to attention: Improving the performance of convolutional neural networks via attention transfer.
\newblock {\em arXiv preprint arXiv:1612.03928}, 2016.

\bibitem{zhao2022decoupled}
B.~Zhao, Q.~Cui, R.~Song, Y.~Qiu, and J.~Liang.
\newblock Decoupled knowledge distillation.
\newblock In {\em Proceedings of the IEEE/CVF Conference on computer vision and pattern recognition}, pages 11953--11962, 2022.

\bibitem{zhao2023embedding}
X.~Zhao, M.~Wang, X.~Zhao, J.~Li, S.~Zhou, D.~Yin, Q.~Li, J.~Tang, and R.~Guo.
\newblock Embedding in recommender systems: A survey.
\newblock {\em arXiv preprint arXiv:2310.18608}, 2023.

\end{thebibliography}

 \vspace{-7.5cm}
\begin{IEEEbiography}[{\includegraphics[width=1in,height=1.25in,clip,keepaspectratio]{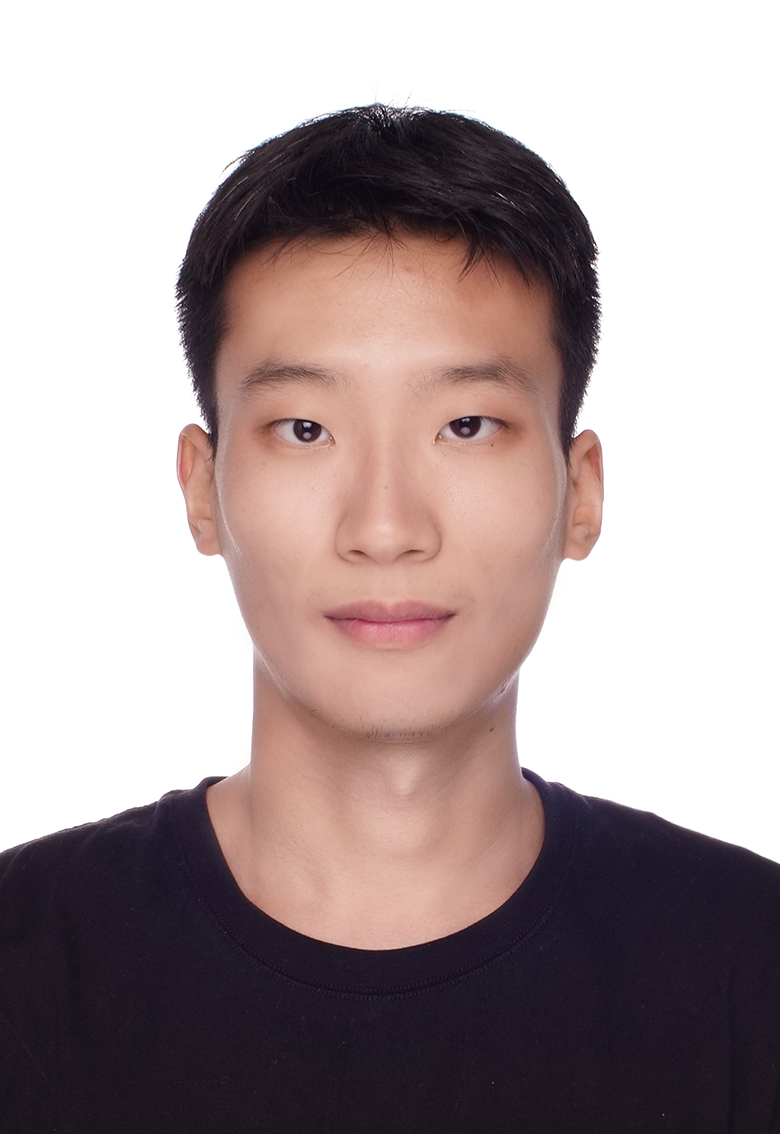}}]{Zhangchi Zhu} received the B.S. degree in Computer Science and Technology from East China Normal University, China, in 2022. He is currently pursuing the master's degree with the Department of Computer Science and Technology, East China Normal University, China. His research interests include recommendation systems, model compression, and large language models.
\end{IEEEbiography}

 \vspace{-7.5cm}
\begin{IEEEbiography}[{\includegraphics[width=1in,height=1.25in,clip,keepaspectratio]{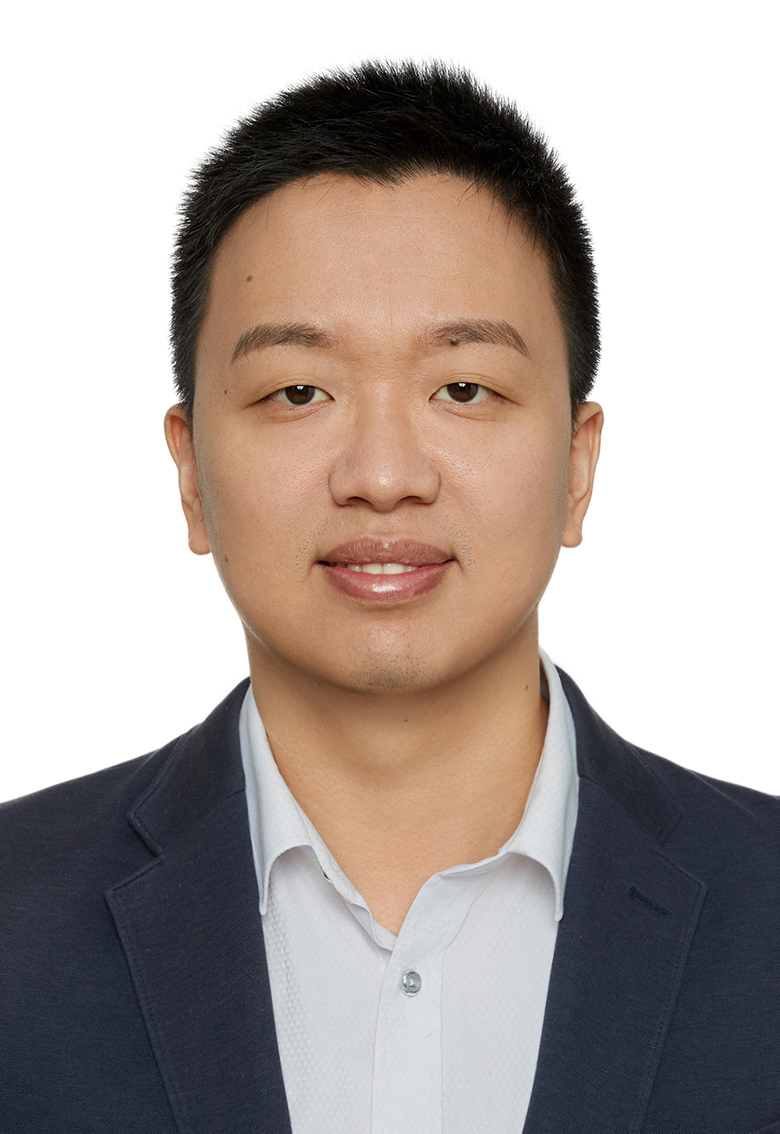}}]{Wei Zhang} received his Ph.D. degree in computer science and technology from Tsinghua university, Beijing, China, in 2016. He is currently a professor in the School of Computer Science and Technology, East China Normal University, Shanghai, China.
His research interests span data mining and machine learning.
his current interests focus on deep recommender systems, interpretable machine learning, large language models, and AI for education.
He is a senior member of China Computer Federation.
\end{IEEEbiography}

\end{document}



\vspace*{-3cm}
\title{Supplementary Material}
\maketitle

\section{Pseudo-code for the Overall Training Process}

\begin{algorithm}[H]
    \caption{Training process of \ourmethod}\label{alg}
    \SetKwInOut{Input}{Input}
    \SetKwInOut{Output}{Output}
    \Input{Training data $\mathcal{R}$, Pretrained teacher model $T$, regularization type $type$, sampling size $Q$, the number of total epochs $E$, rank updating period $K$}
    \Output{Student model $S$.}
    Randomly initialize the Student model $S$.\\
    \For{$epoch=0,1,\cdots,(E-1)$}{
        \uIf{$epoch\%K==0$}{
        Use the student model to predict the preference score of all items and update the rankings.
        }
        \For{each batch $\mathcal{B}\in \mathcal{R}$}{
            Compute $\mathcal{L}_{Base}$ according to Eq.(1).\\
            Compute $\mathcal{L}_{DE}$.\\
            \For{each user $u\in \mathcal{B}$}{
                \uIf(\tcc*[f]{Sampling for pair-wise \ourmethod}){$type$=="pair-wise"}{
                        Draw two items $i$ and $j$ independently from $\mathcal{I}$ with probability $p_u(\cdot)$.
                    }
                \uElseIf(\tcc*[f]{Sampling for list-wise \ourmethod}){$type$=="list-wise"}{
                    Construct $\mathcal{Q}_u$ by drawing $Q$ items independently from $\mathcal{I}$ with probability $p_u(\cdot)$.
                }
                \uElse(\tcc*[f]{Sampling for hybrid \ourmethod}){
                    Construct $\mathcal{Q}_u$ by drawing $Q$ items independently from $\mathcal{I}$ with probability $p_u(\cdot)$.\\
                    Draw two items $i$ and $j$ from $\mathcal{I}$ with probability $p_{u,i}(\cdot)$ and $p_{u,j}(\cdot)$, respectively.\\
                }
            }
            \uIf(\tcc*[f]{Comute pair-wise \ourmethod}){$type$=="pair-wise"}{
                        Compute $\mathcal{L}_{PCKD-P}$ according to Eq.(3).\\
                        $\mathcal{L}_{\ourmethod}\leftarrow\mathcal{L}_{PCKD-P}$.
                    }
                \uElseIf(\tcc*[f]{Compute list-wise \ourmethod}){$type$=="pair-wise"}{
                    Compute $\mathcal{L}_{PCKD-L}$ according to Eq.(5).\\
                    $\mathcal{L}_{\ourmethod}\leftarrow\mathcal{L}_{PCKD-L}$.
                }
                \uElse(\tcc*[f]{Compute hybrid \ourmethod}){
                    Compute $\mathcal{L}_{PCKD-P}'$ according to Eq.(3).\\
                    Compute $\mathcal{L}_{PCKD-L}$ according to Eq.(5).\\
                    Compute $\mathcal{L}_{PCKD-H}$ according to Eq.(8).\\
                    $\mathcal{L}_{\ourmethod}\leftarrow\mathcal{L}_{PCKD-H}$.
                }
            Compute $\mathcal{L}=\mathcal{L}_{base} + \lambda_{DE}\cdot \mathcal{L}_{DE}+\lambda_{\ourmethod}\cdot\mathcal{L}_{\ourmethod}$.\\
            Update the Student model and projectors by minimizing $\mathcal{L}$.
        }
    }
\end{algorithm}

\bigskip
\bigskip
\bigskip
\section{Inference Efficiency}

\begin{table}[htbp]
    \caption{Inference efficiency of \ourmethod with different student model sizes. $\phi$ denotes the ratio of the student's dimensionality to the teacher's. Time (seconds) indicates the inference latency. \#Params denotes the number of parameters. R@10 denotes the ratio of R@10 of \ourmethod to R@10 of the teacher model.}
    \label{tab:effi}
    \centering
    \resizebox{\linewidth}{!}{
    \begin{tabular}{cc|ccc|ccc|ccc} \toprule
        \multirow{2}{*}{Backbone} & \multirow{2}{*}{$\phi$} & \multicolumn{3}{c|}{CiteULike} & \multicolumn{3}{c|}{Gowalla} & \multicolumn{3}{c}{Yelp}\\
         & & Time (s) & \#Params. & R@10 & Time (s) & \#Params. & R@10 & Time (s) & \#Params. & R@10\\
        \midrule
        \multirow{3}{*}{BPRMF} & 1.0 & 23.79 & 11.60M & 1.11 & 125.52 & 20.27M & 1.14 & 172.13 & 19.54M & 1.16\\
         & 0.5 & 17.66 & 5.80M & 1.07 & 97.46 & 10.13M & 1.08 & 144.31& 9.76M & 1.10\\
         & 0.1 & 4.37 & 1.16M & 0.93 & 31.22 & 2.03M & 0.97 & 47.92 & 1.95M & 1.00\\
        \midrule
        \midrule
        \multirow{3}{*}{LightGCN} & 1.0 & 41.36 & 57.98M & 1.08 & 287.62 & 135.11M & 1.13 & 298.48 & 130.30M & 1.11\\
         & 0.1 & 27.20& 5.80M & 0.92 & 129.37 & 13.51M & 1.02 & 155.82 & 13.03M & 1.00\\
         & 0.01 & 9.13 & 0.58M & 0.80 & 61.43 & 1.35M & 0.94 & 79.73 & 1.30M & 0.84\\
        \midrule
        \midrule
        \multirow{3}{*}{SimpleX} & 1.0 & 30.29 & 14.73M & 1.15 & 169.44 & 68.51M & 1.13 & 192.76 & 32.81M & 1.12\\
         & 0.5 & 22.41 & 7.31M & 1.07 & 127.82 & 34.02M & 1.04& 148.97 & 16.35M & 1.03\\
         & 0.1 & 7.81 & 1.45M & 0.98 & 49.38 & 6.77M & 0.98 & 58.29 & 3.26M & 0.95\\
        \midrule
        \midrule
        \multirow{3}{*}{JGCF} & 1.0 & 35.66 & 28.99M & 1.07 & 207.43 & 67.56M & 1.10 & 299.26 & 130.30M & 1.14\\
         & 0.1 & 24.23 & 2.90M & 0.96 & 127.97 & 6.75M & 1.00 & 155.47 & 13.03M & 1.03\\
         & 0.01 & 7.99 & 0.29M & 0.83 & 58.83 & 0.67M & 0.90 & 78.92 & 1.30M & 0.87\\
        \bottomrule
    \end{tabular}
    }
\end{table}
In this section, we report the inference efficiency of our method and comparison methods. All results are obtained by testing with PyTorch on GeForce RTX 3090 GPU. 

To investigate the inference efficiency of our method, we use our method to train students of different sizes. Then, we report the inference latency of these students, together with their recommendation performance.

From the results in Table~\ref{tab:effi}, it is obvious that our method can benefit students of all sizes. Moreover, we achieve comparable performance to the teacher with only 10\% of the number of parameters and about a quarter of the inference latency. When the student is half the size of the teacher, the student can slightly outperform the teacher with only half the storage cost and greatly increase the inference efficiency. Moreover, with the same number of parameters, our method significantly outperforms the teacher, demonstrating that \ourmethod is also effective in improving existing recommendation models.